\documentclass[conference]{IEEEtran}
\IEEEoverridecommandlockouts

\usepackage{subfigure}
\usepackage{cite}
\usepackage{amsmath,amssymb,amsfonts}
\usepackage{algpseudocode}
\usepackage{algorithm}
\usepackage{graphicx}
\usepackage{float}
\usepackage{textcomp}
\usepackage{xcolor}
\usepackage{bm}
\usepackage{multirow}
\def\BibTeX{{\rm B\kern-.05em{\sc i\kern-.025em b}\kern-.08em
T\kern-.1667em\lower.7ex\hbox{E}\kern-.125emX}}
\begin{document}

\title{Fast and Secure Key Generation with Channel Obfuscation in Slowly Varying Environments}

 \author{
 \IEEEauthorblockN{Guyue Li, Haiyu Yang}
 \IEEEauthorblockA{
 Dept. of CSE, SEU \\
 guyuelee, hy\_yang@seu.edu.cn}
 \and
 \IEEEauthorblockN{Junqing Zhang}
 \IEEEauthorblockA{
 Dept. of EEE, University of Liverpool \\
 junqing.zhang@liverpool.ac.uk}
 \and
 \IEEEauthorblockN{Hongbo Liu}
 \IEEEauthorblockA{
 Dept. of CS, UESTC \\
 hongbo.liu@uestc.edu.cn.}
 \and
 \IEEEauthorblockN{Aiqun Hu}
 \IEEEauthorblockA{
 NCRLAB,SEU, \\
 aqhu@seu.edu.cn}
 }

\maketitle

\begin{abstract}
The physical-layer secret key generation has emerged as a promising solution for establishing cryptographic keys by leveraging reciprocal and time-varying wireless channels. However, existing approaches suffer from low key generation rates and vulnerabilities under various attacks in slowly varying environments. We propose a new physical-layer secret key generation approach with channel obfuscation, which improves the dynamic property of channel parameters based on random filtering and random antenna scheduling. Our approach makes one party obfuscate the channel to allow the legitimate party to obtain similar dynamic channel parameters yet prevents a third party from inferring the obfuscation information. Our approach allows more random bits to be extracted from the obfuscated channel parameters by a joint design of the K-L transform and adaptive quantization. A testbed implementation shows that our approach, compared to the existing ones that we evaluate, performs the best in generating high entropy bits at a fast rate and a high-security level in slowly varying environments.
Specifically, our approach can achieve a significantly faster secret bit generation rate at about $67$ bit/pkt, and the key sequences can pass the randomness tests of the NIST test suite. 
\end{abstract}

\section{Introduction}
Physical-layer secret key generation (PKG) has emerged as a promising solution to enable two wireless nodes to generate shared secret keys from the observation and processing of radio channel parameters~\cite{ZHANG2020Frontier}.
Since PKG avoids the need for key distribution by leveraging the reciprocity nature of wireless channels, it is thus appealing to complement traditional cryptographic approaches for scenarios where pre-shared keys may not exist~\cite{li2019physical}.

Existing PKG approaches, however, heavily rely on channel variations and thus suffer from low key generation rate~\cite{9328938} and vulnerabilities under attacks~\cite{Jana} in slowly varying environments.
When the wireless channel varies over time, the keys can be renewed dynamically by sampling the channel parameters.
The channel sampling rate is suggested to be in the order of the maximum Doppler frequency~\cite{mathur08}. However, when users have low mobility, e.g., in a wireless sensor network, the Doppler frequency is low, it requires a long time to establish long enough keys.
Furthermore, the mobility of users and the dynamic change of surrounding objects are often unknown and the maximum Doppler frequency.
For these reasons, there always exists inevitable and unknown temporal correlations between adjacent channel samples, resulting in a large proportion of repeated bit segments. However, these bits segments are scrambled through some permutation or interleaving techniques~\cite{KEEPJ,2016Instant,Bipartite}, the security of the key is still compromised as the permutation information is public.
Another idea of introducing help devices, e.g., relays~\cite{Aldaghri_2020,6716049} and reconfigurable intelligent surface (RIS)~\cite{Low-entropy} to assist secret key generation may improve the key rate and randomness, however, encounters some practical problems, such as the unavailability of trust relays and additional hardware overhead of RIS devices.
Therefore, a fast and secure solution is still required to facilitate the practical usage of PKG in slowly varying environments.

This paper designs a new physical layer key generation approach that can construct fast-changing channel parameters and thus works for slowly varying environments. We introduce a \textbf{\emph{channel obfuscation}} technique, which makes one party obfuscate the channel in a manner that still allows the legitimate party to obtain dynamic channel parameters, yet prevents a third party from inferring the obfuscation information. We show that the channel obfuscation-based secret key generation (CO-SKG) method achieves fast key extraction and a high-security level, even in slowly varying environments.
\begin{figure}
\centerline{\includegraphics[scale = 0.38]{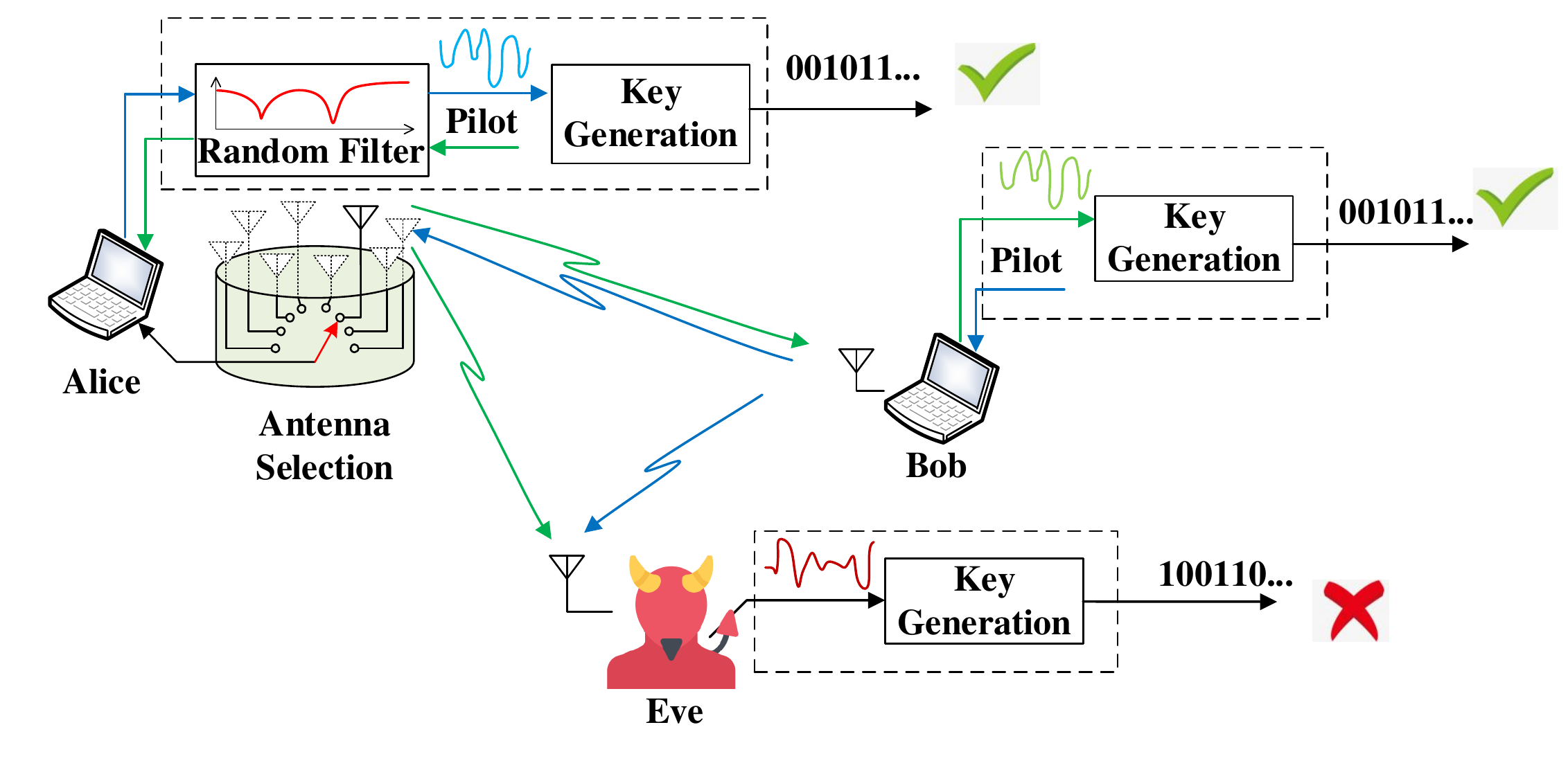}}
\caption{The basic idea of channel obfuscation based secret key generation.}
\vspace{0 em}
\label{fig:system}
\end{figure}

The basic idea underlying CO-SKG is illustrated in Fig.~\ref{fig:system}. Alice and Bob are two legitimate devices, and Eve is an eavesdropper. As a general assumption in PKG~\cite{2021Sum}, Eve is assumed to be located at least half wavelength away from Alice and Bob, so its channel variations are assumed to be independent of Alice and Bob. Alice obfuscates the channel parameters by using a random filter and a random antenna selection module. For each round in these repeated bidirectional channel probings, the transmitted pilot signal and received signal of Alice go through the filter and antenna with the same setting to obtain similar channel parameters with Bob. The settings of the antenna and filter are randomly changed to achieve fast key generation. Since these settings are known only by Alice, Eve can hardly crack the secret key.

CO-SKG builds on past work on exploiting multiple-antenna diversity to increase the bit generation rate~\cite{MAKE}. However, past work typically measures the Received Signal Strength Indicator (RSSI) between each antenna pair in a round-robin way, leading to a periodic variation of RSSI in slowly varying environments. In contrast, CO-SKG measures the CSI between antenna pairs randomly to introduce unpredictable fluctuations in CSI values. CO-SKG also addresses the following practical challenges in using channel obfuscation for secret key generation. First, although the obfuscation information is not public, it might be speculated by a clever Eve from its prolonged channel observations. For instance, Eve may guess the order of antenna pairs by matching its current channel measurement with previous records.
Therefore, a sophisticated design of the channel obfuscation function should be developed to prevent the attacker from inferring the obfuscation information. Second, the reciprocity of the obfuscated CSI is far from satisfactory, and adjacent CSI samples still have some auto-correlation, which may lead to long $0$s and long $1$s in the quantized bit sequences. Therefore, a practical key generation approach must satisfy the keys' agreement, rate, and randomness requirements.

The main contributions of this work are listed as follows:
\begin{itemize}
\item We propose a novel channel obfuscation approach conducive to fast secret bit extraction in slowly varying environments. Meanwhile, our approach prevents a third party from inferring the obfuscation information.
\item We propose an effective key generation approach based on a joint-design of K-L transform and adaptive quantization that would significantly improve the key agreement and randomness.
\item We implemented CO-SKG using Universal Software Radio Peripheral (USRP) software defined radio (SDR) platforms and realizing antenna scheduling with an SP8T switch. Extensive experiments have been conducted, and the results demonstrate that compared with existing typical approaches, CO-SKG can provide higher key agreement, faster key generation rate, and more substantial randomness in three typical, slowly varying scenarios.
\item We have verified that CO-SKG is resilient to various attacks identified as harmful to existing approaches, including the predictable channel attack and position replay attack from active attackers and effective brute-force attack and order speculation attack of passive attackers.
\end{itemize}

\section{Related Works}
There have been ongoing research efforts on fast key generation from wireless channels in slowly varying environments. 
First, diversity techniques, e.g., OFDM and MIMO, have been exploited to improve KGR by extracting more bits from one channel sample~\cite{QinExploiting,Applying_12TIFS,5934929}. For example, Liu~\textit{et al}.~\cite{CGC} use channel response from multiple OFDM subcarriers to provide fine-grained channel information, and Zeng~\textit{et al}.~\cite{MAKE} use multiple-antenna diversity to increases BGR by more than four times over single-antenna systems.
Despite these efforts, these protocols still require dynamic environments to meet the requirement of randomness and to refresh keys. To deal with it, opportunistic beamforming~\cite{Applying_12TIFS,6567033,8543098} is exploited in multiple antenna systems. However, its performance has only been evaluated through theoretical analysis and simulations~\cite{liguyue2017}.

Second, secret key generation with helper devices such as relays has also been proposed~\cite{Aldaghri_2020,6716049,6567033}.
Although this relay-based solution may increase the key rate, it is affected by design challenges. The key rate depends on the relay movement, but it is impractical to persuade a relay to move all the time. Moreover, relays are not always available and trustable~\cite{Gui2015Untrusted}, which limits the practical usage of this solution. Recently, Paul~\textit{et al}~\cite{Low-entropy} has reported promising experimental results on integrating a passive reconfigurable intelligent surface (RIS) device to help in achieving a higher secret bit generation rate in static environments. The RIS device does not have a trustworthy problem. However, it adds additional hardware overhead to the key generation system.

Another solution is to apply permutation~\cite{Bipartite,2016Instant} or interleaving technique~\cite{KEEPJ,9201305} to increase the randomness of the weak key generated in the static environments.
The channel samples are rearranged according to a pseudo-random sequence generated by specific algorithms. However, for legitimate users to agree on the same key, the pseudo-random sequence of interleaving is shared on the public channel, which means that an attacker also knows the information. She can guess the original weak key and then interleave it for correctness verification. Hence, from the view of key cracking, it is futile to interleave the CSI samples.

Our work differs from existing works in the following significant ways.
First, we perform extensive real-world measurements in various slowly varying environments and settings to determine the effectiveness and security of the proposed CO-SKG approach.
Second, we propose a channel obfuscation method to allow Alice and Bob to obtain similar dynamic channel parameters yet prevent Eve from inferring the obfuscation information.
Third, we further increase the secret bit rate and randomness by a joint design of K-L transform and adaptive quantization.

\section{Problem Formulation and the CO-SKG Scheme}
\subsection{Problem Formulation}
As shown in Fig.~\ref{fig:system}, Alice (A) and Bob (B) intend to generate a shared secret key from their reciprocal channel state information (CSI) over $N$ orthogonal frequency division multiplexing (OFDM) subcarriers. Eve (E) attempts to find the key based on its channel observations and information transmitted over the public channel.
Alice is equipped with $M$ antennas, and Bob has a single antenna.
They share $M$ spatial channels, which are collected one after another,
by switching the working antenna of Alice.
Therefore, the system works in a single-input single-output (SISO) mode, and Eve is assumed to be equipped with a single antenna.


Alice and Bob perform $K$ periods of channel probing to obtain their CSI samples to generate secret keys.
Under a slowly varying environment, the wireless channel changes little during these $K$ periods.
For the $k$-th round probing, the signal received on subcarrier $n$ by user $u \in \{A, B, E\}$ can be expressed as
\begin{align}
Y_u (k,n) = H_u(m_k,n) S(k,n) + Z_u(k,n),
\end{align}
where ${H_u(m_k,n)}$ represents the spatial channel of the $m_k$-th antenna in the $k$-th round, $S(k,n)$ represents the known probe signal, and $ Z_u(k,n)$ is the noise.
In the existing work of \cite{MAKE}, the antennas are used in a round-robin way, which implies that
\begin{align}\label{eq:mk}
m_k = (k+c)_M,
\end{align}
where $(\cdot)_M$ represents the modulo $M$ operation and $c$ is a constant integer with the range of $\{1,2,\cdots, M-1\}$.
Based on the received signal, the estimated CSI at user $u$ is given by
\begin{align}\label{eq:2}
{\hat H}_u(k,n) = H_u(m_k,n) + Z'_u(k,n),
\end{align}
where $Z'_u(k,n)$ represents the equivalent noise
after estimating $H_u(m_k,n)$.
According to the principle of channel reciprocity,
when Alice and Bob probe the channel within the channel coherence time, the estimated CSI values, i.e., ${\hat H}_A(k,n)$ and ${\hat H}_B(k,n)$ should be highly correlated, which allows Alice and Bob to extract the same secret bit sequence from their CSI estimates, respectively. The secret bit sequence is continually renewed in the following key generation time rounds.

\begin{figure}
\subfigure[Single antenna]{
\includegraphics[scale = 0.275]{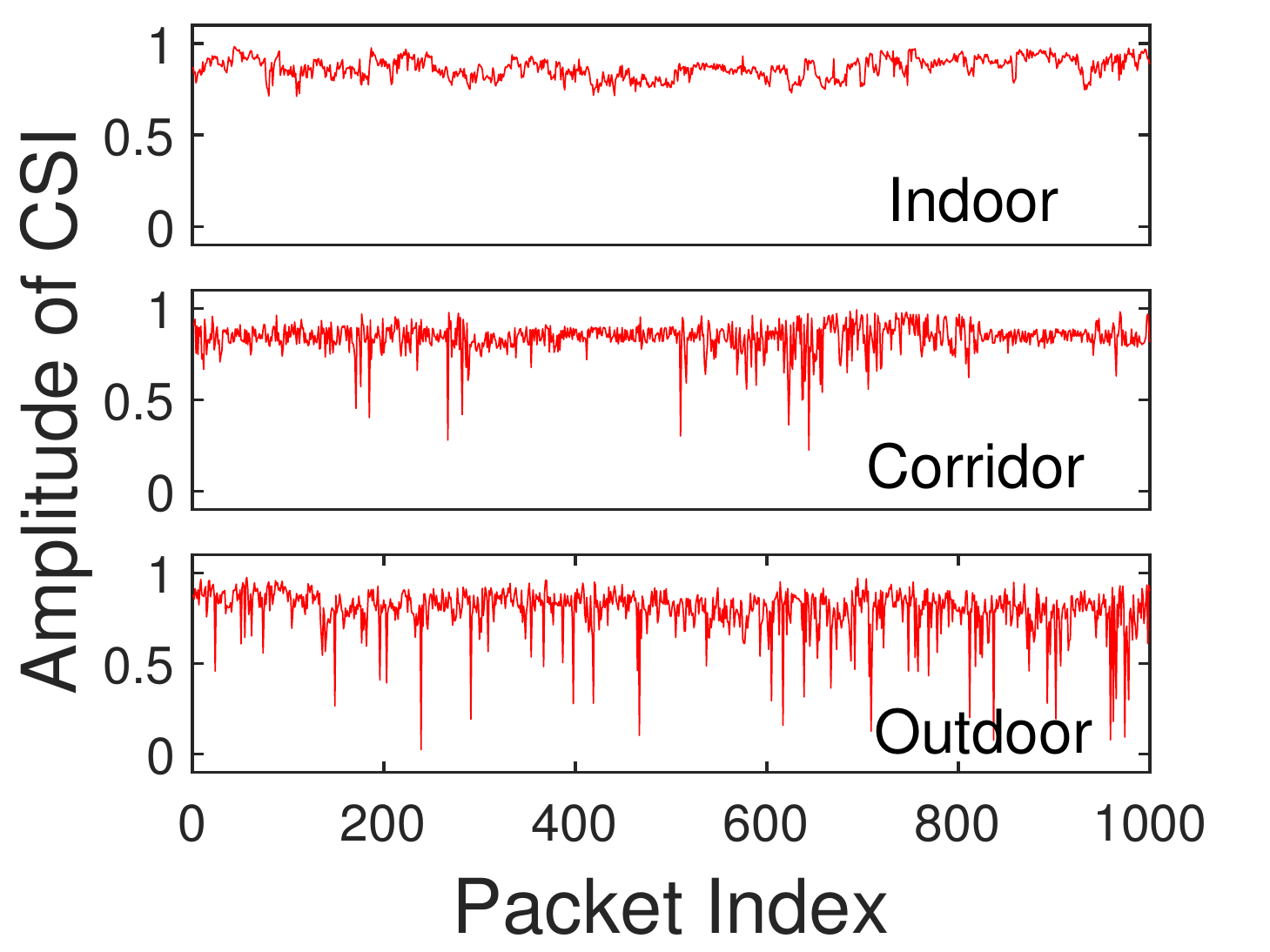}
\label{fig:csi_vary}
}
\subfigure[Multiple antennas]{
\includegraphics[scale = 0.275]{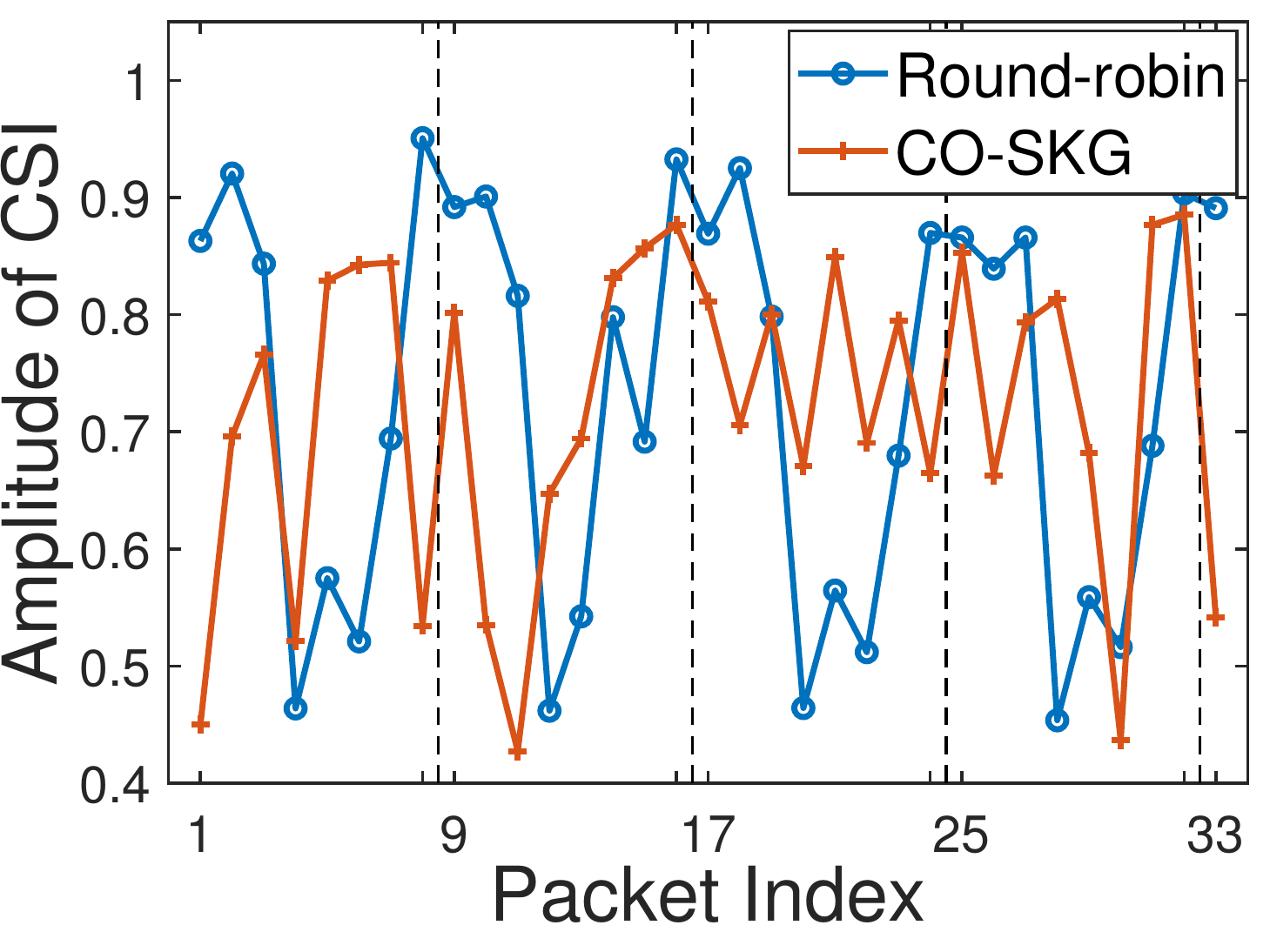}
\label{fig:csi_r}
}
\vspace{-0.5em}
\caption{The amplitude of CSI collected in slowly varying environments. }
\vspace{-1em}
\end{figure}

However, the CSI for each antenna pair may have little fluctuation in slowly varying environments. When antennas are scheduled in a fixed order periodically, we have
\begin{align}\label{eq:4}
{\hat H}_u(k+M,n) \approx {\hat H}_u(k,n),
\end{align}
which indicates that the collected CSI samples may change regularly with time.
To verify it, We conducted experiments under three slowly varying scenarios, i.e.,
indoor, corridor, and outdoor for validation. Two USRP N210 devices operating in the $2.535$ GHz channel are deployed as Alice and Bob, which remain stationary during the experiments.
We collected CSI samples in $1000$ time rounds, with a time interval of $3$ seconds.
Fig.~\ref{fig:csi_vary} shows that for a fixed antenna pair, the amplitude of CSI on one subcarrier has slight variation during the $1000$ rounds.
Next, we let Alice connect to an SP8T switch to realize the function of antenna selection. When the antenna is used in a round-robin way, as predicted in (\ref{eq:4}), the blue curve in Fig~\ref{fig:csi_r} exhibits significant regularity with a period of $8$. These periodic changes in CSI may result in a large proportion of repeated bit segments in the quantization results.

\subsection{The CO-SKG Scheme}
\subsubsection{Preliminary Study}
One way to realize channel obfuscation is by randomly changing the index of the employed antenna in each packet.
In this way, the order of CSI samples varies from one probing round to another. As a result, the CSI samples will have more fluctuation,
largely avoiding the periodic repetition.
We also experiment with randomly changing antennas on the USRP platform, and the amplitude of CSI sequences is shown in Fig.~\ref{fig:csi_r}.
As expected, the CSI estimated in different rounds has significant variation, and the CSI values' regularity has disappeared.
\subsubsection{Practical Challenges}
The above preliminary study provides encouraging results on the feasibility of improving channel fluctuation with channel obfuscation in slowly varying environments.
However, to achieve consistent secret keys with channel obfuscation securely and efficiently, there are still two main challenges to be addressed.

\textbf{\emph {How to hide the obfuscation information?}} A desired channel obfuscation approach should obfuscate the channel without leaking the obfuscation information.
However, when the channel fluctuation is simply caused by antenna switching, the obfuscation information might be known to Eve.
In a slowly varying environment, the CSI of Eve also varies little when Alice selects the same transmit antenna. This fact helps Eve to derive the relationship of the used antenna pair by matching the current CSI with previous CSI records. 

On this basis, a sophisticated design of channel obfuscation should be developed to prevent Eve from knowing the obfuscation information.

\textbf{\emph {How to produce secret keys effectively?}} Converting the obfuscated CSI samples into consistent secret keys is also challenging.
Since the equivalent obfuscation channel changes rapidly,
smoothing algorithm~\cite{7458871}, usually used to improve the similarity of channel parameters in existing works,
does not apply to the obfuscated CSI samples,
resulting in an inconsistent high rate.
When the inconsistent rate of the raw key is high, the information reconciliation imposes a significant burden to correct these errors, which can significantly affect the efficiency of key generation.
Besides, although channel obfuscation improves the fluctuation of CSI samples, there is still inevitable auto-correlation between these CSI samples, which leads to long $0$ and long $1$ in the quantized results.
Therefore, a practical key generation approach needs to satisfy the high requirements of agreement rate and randomness of keys.

\begin{figure}
\centerline{\includegraphics[scale = 0.35]{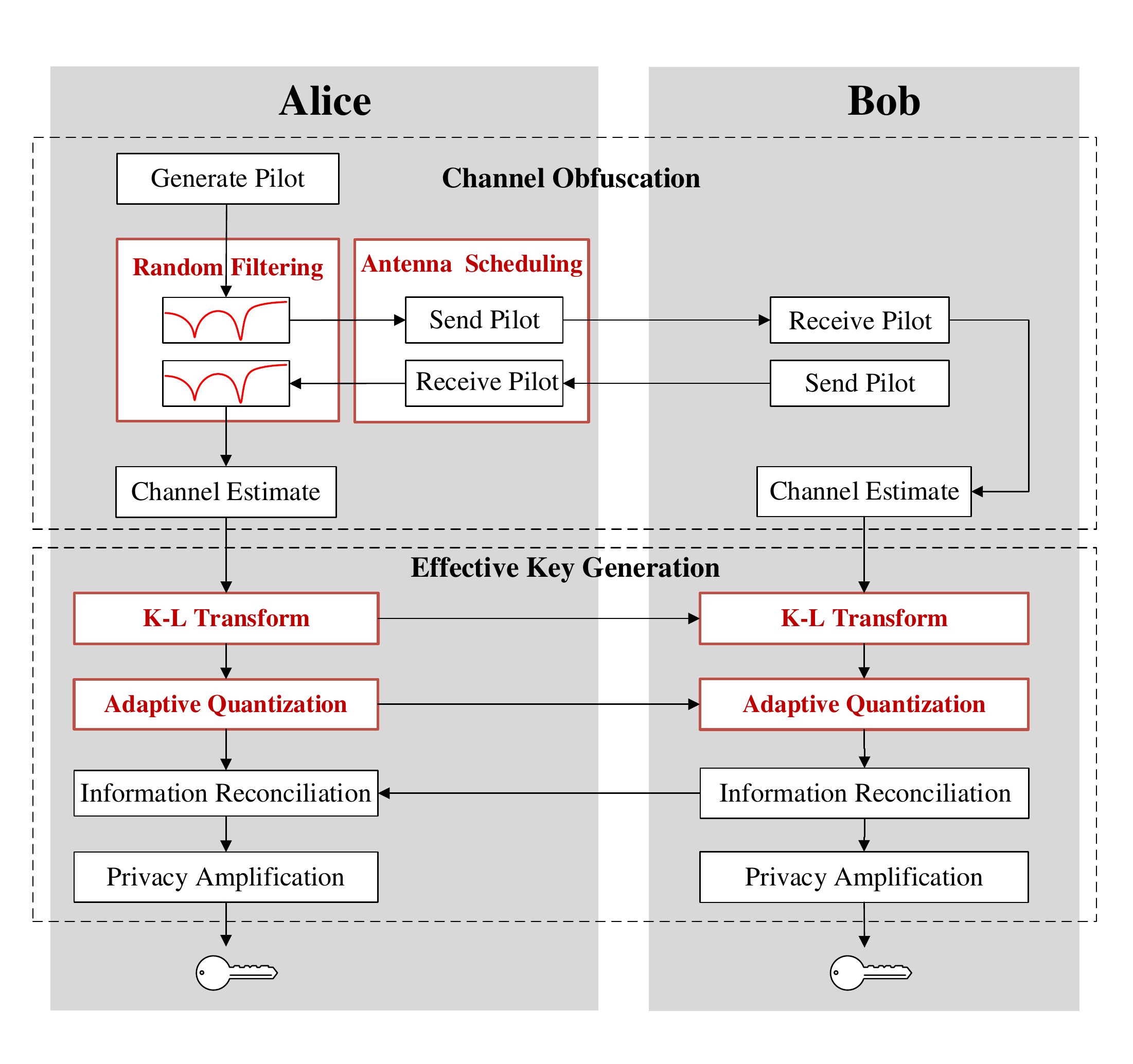}}
\vspace{-1em}
\caption{The flow of CO-SKG.}
\vspace{-1em}
\label{fig:flow}
\end{figure}

To cope with these challenges, we propose a CO-SKG approach, which is divided into two parts, i.e., channel obfuscation and effective key generation, as shown in Fig.~\ref{fig:flow}.
Alice and Bob obtain the channel estimates during channel obfuscation by sending pilot signals to each other in turn. Different from existing works, Alice incorporates the technologies of random filtering and antenna scheduling to fluctuate the CSI samples.
After channel obfuscation, Alice and Bob convert their channel estimates into secret keys through effective key generation.
We exploit K-L transform and adaptive quantization to obtain raw key bits with the high agreement and strong randomness. Then, Alice and Bob obtain consistent keys from their raw key bits through information reconciliation and privacy amplification.
Details of the protocol design of channel obfuscation and effective key generation are elaborated in Section~\ref{sec:rp} and Section~\ref{sec:ekg}, respectively.

\subsection{Attack Model}\label{sec:attack}
We consider four types of attacks, which exploit the vulnerability of key generation in slowly varying environments. In the first two types of attacks, Eve intends to infer the key of the current round, while in the last two types of attacks, Eve aims at cracking the renewed key with an assumption that the current key has been known.

\textbf{\emph{Predictable Channel Attack}}~\cite{Jana2009effectiveness}: Eve tries to cause desired or predictable changes in the channel measurements by being the one who moves between Alice and Bob.

\textbf{\emph{Position Replay Attack}}~\cite{Applying_12TIFS}: In stationary environments, the channel parameters might be identical at the same position even when the measurements are made at different times. In this case, Eve will record the position of Bob and moves to it after Bob leaving to acquire similar channel measurements.

\textbf{\emph{Effective Brute-force Attack}}: In slow-varying environments, the successive key generation processes will, with high likelihood, merely result in the production of a large proportion of repeated bit segments. Eve can predict the bit change rate of legitimate users according to the variation of its channel observations. Hence, the time-complexity of brute-force to the renewed key can be primarily shortened as Eve only needs to try a small range of bits that have been changed.

\textbf{\emph{Order Speculation Attack}}:
In the PKG approaches based on permutation, the fluctuation of CSI samples is caused by the randomness of the sample order. A clever Eve will speculate the order from its channel observations and the information transmitted over the public channel. When the order is known, the process of key cracking returns to that in the effective brute-force attack.

Besides, Eve is assumed to be more interested in the
extracted keys between Alice and Bob but not disrupting
their key establishment process. Therefore, we also assume
that Eve can neither jam the communication channel between Alice and Bob nor modify the information exchanged between them.

\section{Channel Obfuscation}
\label{sec:rp}
\subsection{Design Philosophy}\label{sec:codp}
The desired CSI $\hat{H}_u(k,n)$ should meet the following three requirements.
\begin{itemize}
\item {Randomness:} $\hat{H}_A(k,n)$ should change over $k$ in an unpredictable way. As a counterexample, periodic variable $m_k$ in (\ref{eq:mk}) should be avoided, because it obfuscates the CSI samples with regular changes that can be predicted.
\item {Reciprocity:} Channel obfuscation should not affect the reciprocity principle, i.e., Alice and Bob are able to obtain similar CSI samples as $\hat{H}_A(k,n) \approx \hat{H}_B(k,n)$.
\item {Security:} The obfuscation information should not been leaked from $\hat{H}_E(k,n)$ to Eve, otherwise the security of $\hat{H}_A(k,n) $ is compromised.
\end{itemize}
Accordingly, we design a time-varying function, ${\bf G}(k,n)$, which is given by
\begin{align}
{\bf G}(k,n) = \alpha(k,n) {\bf g}_k,
\end{align}
where ${\bf g}_k\in \mathbb{R}^{M\times 1}$ is the antenna selecting vector,
in which the $m_k$-th element is $1$ and others are $0$, and
$\alpha(k,n)$ is the weighting coefficient of the selected channel.
Denoting the channel vector as $ {{\bf H}}_u(n) = [{H}_u(1,n),
{H}_u(2,n),\cdots,{H}_u(M,n)]$,
the CSI estimated by user $u$ at the $k$-th round is
\begin{align} \label{eq:gkn}
\hat{H}_u(k,n) &= {{\bf H}}_u(n) {\bf G}(k,n) + Z'_u(k,n) \nonumber\\
&=\alpha(k,n) H_u(m_k,n) + Z'_u(k,n).
\end{align}
We incorporate random variables $\alpha(k,n)$ and $m_k$ into function ${\bf G}(k,n)$ to obtain the desired CSI. The necessity of their combination is expounded as follows.

\textbf{{Case 1:}} When $\alpha(k,n)=1$ and $m_k$ is an integer uniformly distributed variable, $m_k \sim U[1, M]$, the CSI is obfuscated by randomizing index $m_k$ of $M$ spacial channels. The obfuscated CSI can meet the requirements of randomness and reciprocity.
However, its variation range is limited mainly by the number of antennas $M$.
When Alice uses the same antenna $m_k$ for channel probing round of $k$ and $k'$, Eve is able to observe $\hat{H}_E(k,n) \approx \hat{H}_E(k',n)$. With the increase of channel probing time $K$, the CSI differences between different antenna pairs will become more distinguishable.

\textbf{{Case 2:}} When $m_k =(k)_M$ and $\alpha(k,n)$ is a complex random variable, the CSI is obfuscated by multiplying the periodic CSI with a random coefficient $\alpha(k,n)$ that is known only by Alice.
However, when only random coefficient is exploited, Eve is possible to speculate the obfuscation information of the coefficient according to the known periodic index of $m_k$. To clarify the speculation process, we omit the noise term and put the CSI with the same $m_k$ into a vector. The CSI vectors of Alice and Eve are respectively given by
\begin{align}
\vec H_{A,n}&=\left[\alpha(k,n),\alpha(k+M,n),\cdots\right] H_A\left((k)_M,n\right),\\
\vec H_{E,n}&=\left[\alpha(k,n),\alpha(k+M,n),\cdots\right] {H_E}\left((k)_M,n\right).
\end{align}
It is worth noting that Eve can hardly know the exact value of $\vec H_{A,n}$, due to the unknown information of $ H_A((k)_M,n)$ and $ {H_E}((k)_M,n)$, but she can observe the obfuscation information of coefficients. For example, the normalized amplitude of $\vec H_{A,n}$ and $\vec H_{E,n}$ are the same.

Notably, the function ${\bf G}(k,n)$ solely relies on random variable $m_k $ or $\alpha(k,n)$ will leak the obfuscation information. Fortunately, from the above analysis, we find that the index $m_k$ and coefficient $\alpha(k,n)$ have mutual remedying parameters in hiding the obfuscation information.
First, $\alpha(k,n)$ prevents Eve from knowing the obfuscation information of $m_k $, as the channel estimates of two-time rounds are different due to the different $\alpha(k,n)$.
Second, $m_k$ also prevents Eve from knowing the obfuscation information of $\alpha(k,n)$, as Eve can hardly collect CSI with the same antenna to find the obfuscation information of coefficients.
Therefore, by incorporating random index $m_k$ and random coefficient $\alpha(k,n)$, the desired CSI can be achieved through the channel obfuscation function in (\ref{eq:gkn}).


\subsection{Design Protocol}
To implement the function ${\bf G}(k,n)$ in reality, we propose a channel obfuscation protocol by adopting the technologies of \emph {random antenna scheduling} and \emph {random filtering}.
The former randomizes the index $m_k$ by randomly selecting the employed antenna of Alice for each channel probing. The latter randomize the value of $\alpha(k,n)$ by adding a time-varying finite impulse response (FIR) filter with $L_f$ taps to process the pilot signal and received signal at Alice. 
The tap coefficients of the FIR filter are denoted by $\vec a = [a_1, a_2, \cdots, a_{L_f}]$, where $a_i \sim \mathcal{CN} (0, 1)$ is a random variable following the complex Gaussian distribution with zero mean and unit variance.

The pilot signal of Alice is cyclic convolved with
the impulse response of the FIR filter as well as the channel.
Since the cyclic convolution in the time domain is equivalent to the multiplication in the frequency domain, the value of $\alpha(k,n)$ in the vector $\alpha(k)=[\alpha(k,1),\alpha(k,2),\cdots,\alpha(k,N)]$
is
\begin{align}\label{eq:alpha}
\alpha(k,n) =\sum\limits_{i = 1}^{L_f} {a_i}{e^{ - j2\pi \frac{n i}{N}}}.
\end{align}
The detailed channel obfuscation for the $k$-th time round of key generation
is summarized in Algorithm 1.

\begin{algorithm}
\caption{Channel obfuscation.}
\label{alg:CO}
\hspace*{\algorithmicindent} \textbf{Input:} The pilot signal $S=[S(1),S(2),\cdots, S(N)]$. \\
\hspace*{\algorithmicindent} \textbf{Output:} The CSI vectors ${{\vec H}_A}({\rm{k}})$ and ${{\vec H}_B}({\rm{k}})$.
\begin{algorithmic}[1]
\State \textbf{Alice:}
\For {$i:= 1$ to $L_f$}
\State $a_i \leftarrow $ random variable following $\mathcal{CN} (0, 1)$.
\EndFor
\State Generate $\alpha(k)$ with $\{a_1, a_2, \cdots, a_{L_f}\}$ according to~(\ref{eq:alpha}).
\State $S_A \leftarrow $ element-wise multiple $S$ and $\alpha(k)$.
\State $m_k \leftarrow $ random integer following $U[1, M]$.
\State Sends $S_A$ to Bob via the $m_k$-th antenna.
\State \textbf{Bob:}
\State Receives the signal $Y_B (k,n)$.
\State Estimates CSI by ${{\hat H}_B}(k,n)=\frac{Y_B (k,n)}{S(n)}$ and obtains
\begin{small}${{\vec H}_B}({\rm{k}})={\left[ {{{\hat H}_B}(k,1),{{\hat H}_B}(k,2), \cdots ,{{\hat H}_B}(k,N)} \right]^T}$\end{small}.
\State Sends $S$ to Alice.
\State \textbf{Alice:}
\State Receives signal $Y_A (m_k,n)$ on $m_k$-th antenna .
\State $Y'_A (k,n) \leftarrow$ multiple $Y_A (k,n)$ and $\alpha(k,n)$.
\State Estimates CSI by ${{\hat H}_A}(k,n)=\frac{Y'_A (k,n)}{S(n)}$ and obtains
\begin{small}${{\vec H}_A}({\rm{k}})={\left[ {{{\hat H}_A}(k,1),{{\hat H}_A}(k,2), \cdots ,{{\hat H}_A}(k,N)} \right]^T}$\end{small}.
\end{algorithmic}
\end{algorithm}

Thanks to the reciprocity principle, the CSI vectors received by Alice and Bob, i.e., ${{\vec H}_A}({\rm{k}})$ and ${{\vec H}_B}({\rm{k}})$, are proportional to each other.
In the following rounds of channel probing, the process will be repeated with changing filter coefficients and antenna index.
After $K$ time rounds, Alice and Bob obtain their CSI matrices of \begin{small}${{{\bf{\hat H}}}_A} = \left[ {{{\vec H}_A}(1),{{\vec H}_A}(2), \cdots ,{{\vec H}_A}(K)} \right]$\end{small} and \begin{small}${{{\bf{\hat H}}}_B} = \left[ {{{\vec H}_B}(1),{{\vec H}_B}(2), \cdots ,{{\vec H}_B}(K)} \right]$\end{small}, respectively.
Then, they use these observations to generate the secret keys.


\section{Effective Key Generation}\label{sec:ekg}
Alice and Bob will generate secret keys from the collected CSI matrices ${\bf \hat{H}}_A$ and ${\bf \hat{H}}_B$, each of which has $N$ subcarriers and $K$ samples.

\subsection{K-L Transform}
\begin{figure}[!t]
\centerline{\includegraphics[scale = 0.35]{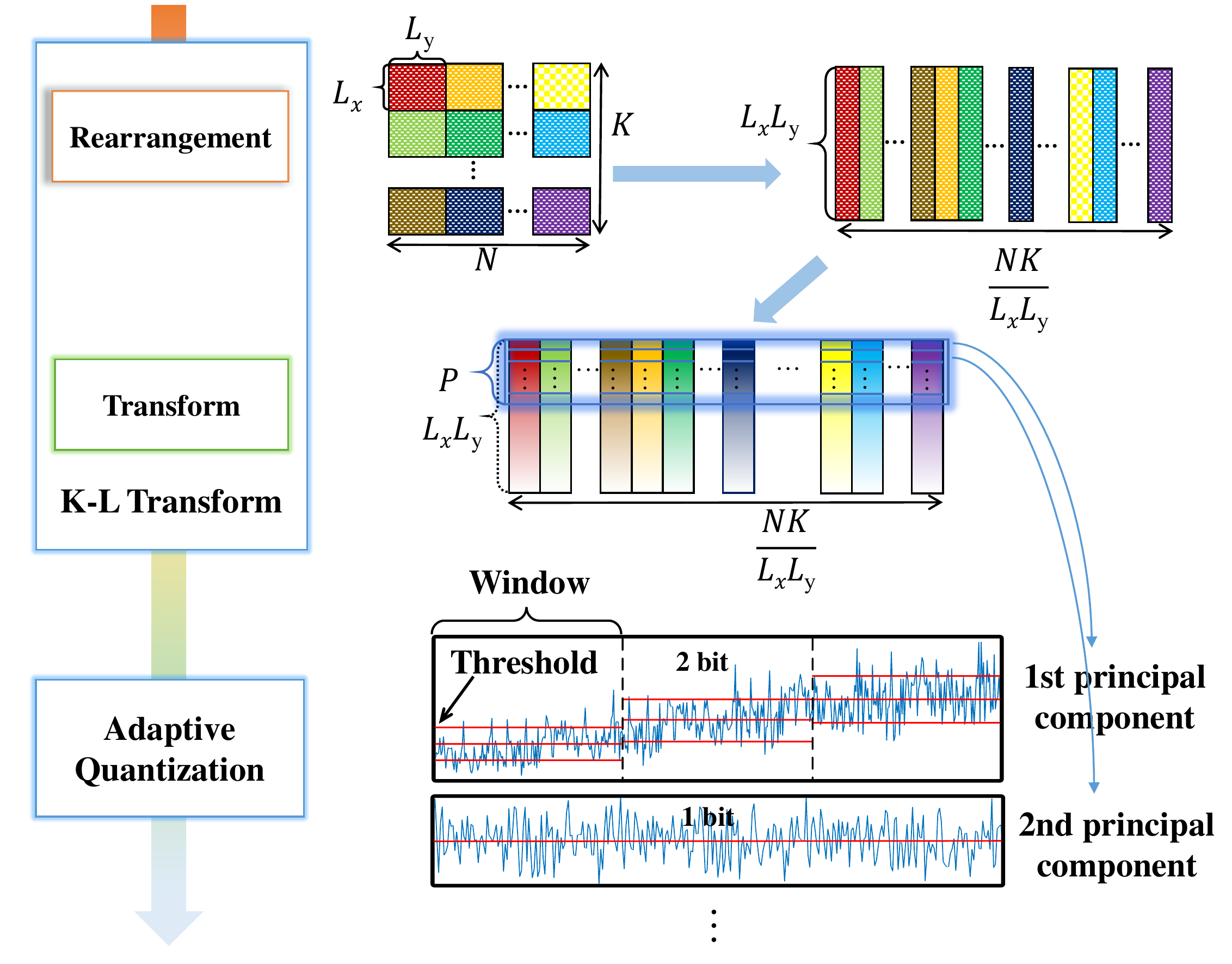}}
\caption{K-L transform and adaptive quantization.}
\label{fig:4}
\end{figure}

We apply the K-L transform to CSI samples for reducing the residual correlation between them. The process has two steps, i.e., CSI rearrangement and transform, as shown in Fig.~\ref{fig:4}.

\textbf {Rearrangement}: As shown in Fig.~\ref{fig:4}, elements in
${\bf \hat{H}}_A$ and ${\bf \hat{H}}_B$ are segmented into multiple blocks, with $L_x \times L_y$ CSI values in each block.
There are $\frac{N}{L_x}$ and $\frac{K}{L_y}$ blocks in the direction of subcarrier and samples, respectively.
The CSI values in different blocks are assumed to be uncorrelated, so we aim at removing the correlation between CSI values in the same block.
Therefore, $L_x $ and $L_y$ are designed according to the assumption. The CSI values in the same block are vectorized into a column. Then, ${\bf \hat{H}}_A$ and ${\bf \hat{H}}_B$ are rearranged as ${\bf \dot H}_A$ and ${\bf \dot{H}}_B$, each has $L_x \times L_y$ rows and $\frac{N}{L_x}\times \frac{K}{L_y}$ columns.

\textbf {Transform}: Based on ${\bf \dot H}_A$, Alice calculates the channel covariance matrix, ${\bf R}_A$, and performs the eigenvalue decomposition of ${\bf R}_A$ to obtain the eigenvector matrix ${\bf U}_A$ and eigenvalue matrix $\Lambda_A$. Only a small part of principal components are suitable for key generation, and others are severely corrupted by noise.
So we construct the transform matrix ${\bf V}_A$ by selecting the first $P$ column vectors in ${\bf U}_A$, which corresponds to $\eta$ percentage of energy.
Alice sends ${\bf V}_A$ to Bob through the public channel and they complete the transform by multiplying ${\bf \dot H}_A$ and ${\bf \dot{H}}_B$ with ${\bf V}_A$ as ${\bf \ddot H}_A = {\bf V}_A {\bf \dot H}_A, {\bf \ddot H}_B= {\bf V}_A {\bf \dot H}_B$, respectively.
The transmission of eigenvector over an insecure public channel can cause information leakage and the information leakage ratio is $\eta_1 \approx 1/(\frac{N}{L_x}\times \frac{K}{L_y})$ as analyzed in~\cite{PCA}.

\subsection{Adaptive Quantization}
The transformed CSI values in ${\bf \ddot H}_A$ and ${\bf \ddot H}_B$ are converted into bit sequences, ${\bf q}_A$ and ${\bf q}_B$, respectively, through an adaptive quantization algorithm that can adjust the \textit{quantization level} and \textit{thresholds} dynamically. These bit sequences are referred to as raw keys.

Since ${\bf \ddot H}_A$ and ${\bf \ddot H}_B$ contain multiple
principal components with different signal-to-noise ratios (SNRs), a fixed quantization level does not apply to all components.
Thus, we employ flexible quantization levels in the quantization
algorithm. The levels $L_p$ are decided by the desired disagreement ratio of raw keys and the corresponding SNRs. 

For each component, the threshold should be adaptive to the variation of the values to avoid long $0$s and long $1$s. Thus, we add windows to the CSI sequence and quantize them in each window.
In the design of the window length, $L_w$, we trade-off between the randomness and the disagreement ratio of the raw key. In the implementation, $L_w$ is set according to the variance of the CSI values in the window. In each window, we apply multiple-level quantization with a guard band. The threshold is set based on probability, and the quantization bits are with Gray code.

Finally, the bit sequences, ${\bf q}_A$ and ${\bf q}_B$, are the combination of quantization results of each column in ${\bf \ddot H}_A$ and ${\bf \ddot H}_B$, respectively.

\subsection{Information Reconciliation and Privacy Amplification}
To obtain identical bit sequences, we apply a BCH code to correct the disagreements in raw keys according to their disagreement ratio. Bob calculates the syndrome of the ${\bf q}_B$ and sends it to Alice over the public channel.
Alice corrects ${\bf q}_A$ according to the received syndrome. 
Transmitting the syndrome of BCH transmitted over public channels leaks information to eavesdroppers.
The leakage rate in information reconciliation is $\eta_2= L_s/L_q$ where $L_q$ and $L_s$ are the lengths of the raw key and the syndrome, respectively.

Privacy amplification allows legitimate users to distill a shorter but almost completely secret key from a common random variable about which Eve has partial information.
We use Message-Digest Algorithm 5 (MD5) for privacy amplification. MD5 is a widely used hash function which maps data of arbitrary size to data of 128 bits.
Considering the information leakage in the phases of K-L transform and information reconciliation, Alice and Bob should at least generate
$L_{req} = \lceil\frac{128}{ (1-\eta_1) (1-\eta_2)}\rceil$
bits common random sequence, where $\lceil \cdot \rceil$ represents ceiling operation.


\section{Performance Evaluation}
\subsection{Experimental Setup}
\begin{figure}
\subfigure[Hardware testbed]{
\includegraphics[scale = 0.13]{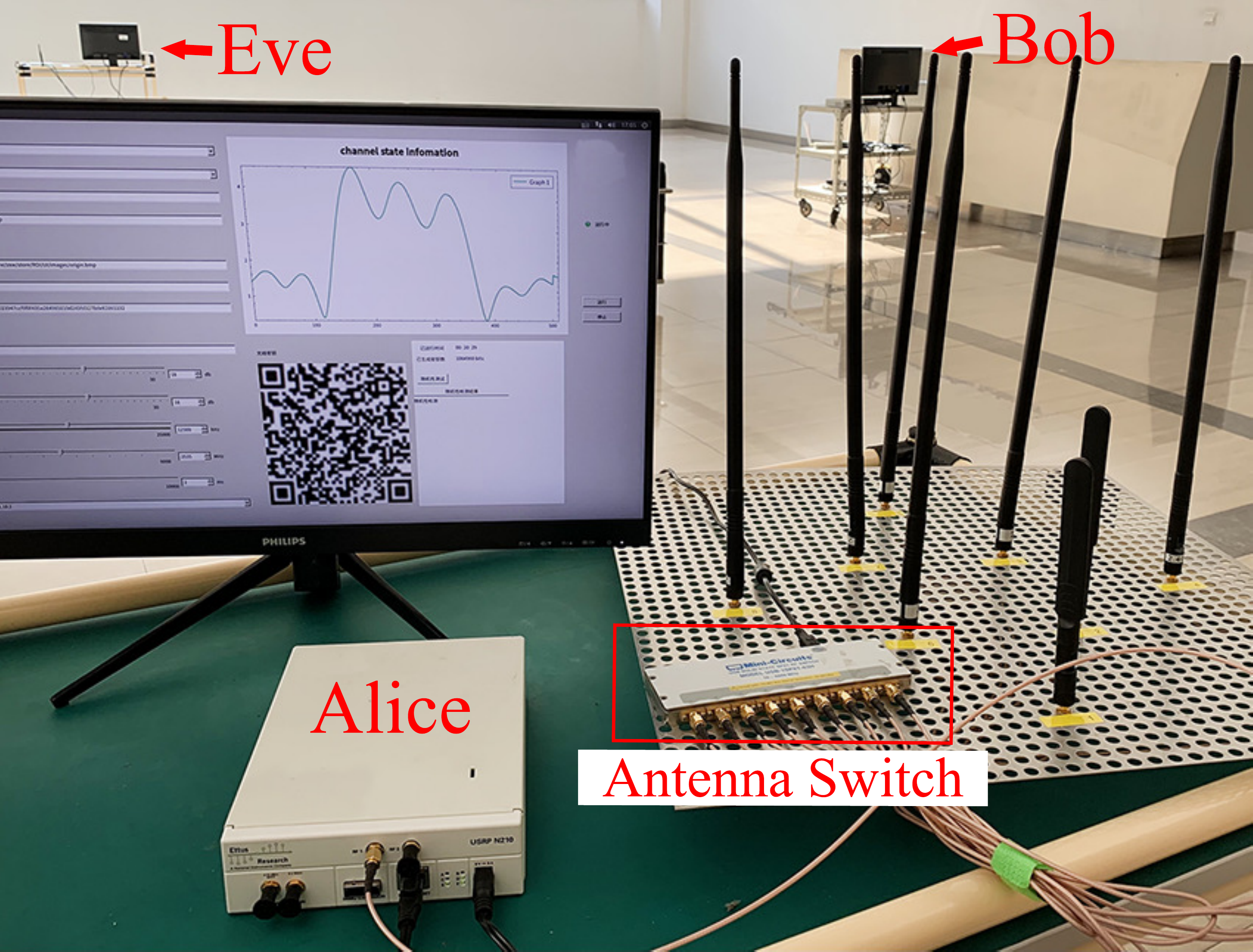}
\label{fig:plfa}}
\subfigure[Antenna switch]{
\includegraphics[scale = 0.25]{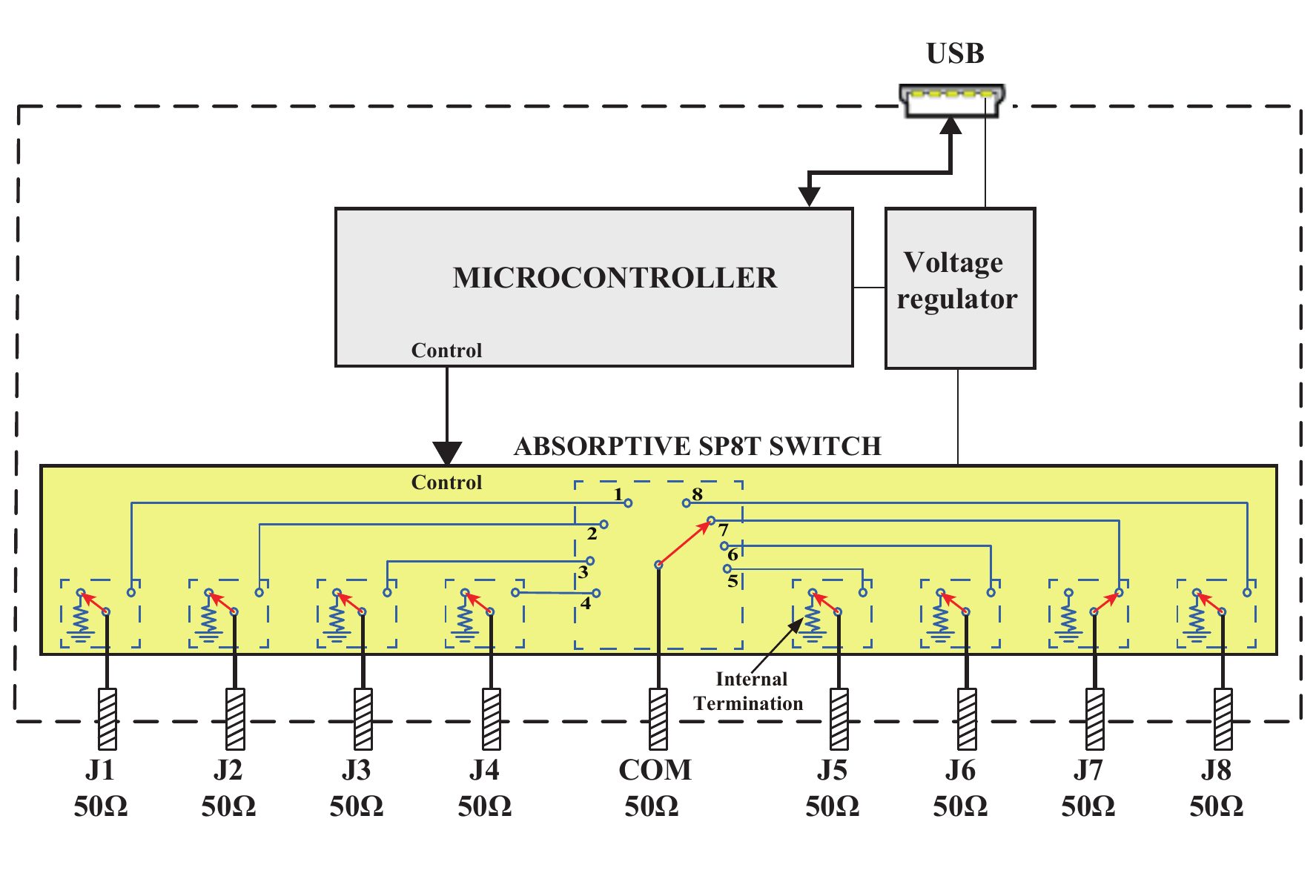}
\label{fig:plfb}}
\caption{The experimental platform.}
\label{fig:plf}
\end{figure}

To evaluate the performance of the proposed CO-SKG approach in real environments, we built a hardware testbed using USRP SDR platform. Three USRP N210 SDR platforms embedded with the CBX daughterboards~\cite{USRP} were used as Alice, Bob, and Eve, as shown in Fig.~\ref{fig:plfa}. The experiments were carried out at the $2.535$ GHz channel. 
The USRP receiver received the signal, which was transferred to the PC and processed by MATLAB.
To enable the function of antenna scheduling in the CO-SKG approach, Alice is connected with an SP8T switch~\cite{Switch} as shown in Fig.~\ref{fig:plfb}. 
Alice sends instructions to the microcontroller through software to control the SP8T switch to connect COM to one of J1 to J8.
We carried out extensive experiments in the Purple Mountain Laboratories, China, under three slowly varying scenarios, i.e.,
indoor, corridor, and outdoor. Three USRP devices remain stationary, but there might be people or vehicles moving around.
We collected at least $1000$ CSI vectors, each containing $N=512$ CSI values on these OFDM subcarriers.

We adopted three evaluation metrics: bit mismatch rate (BMR), bit generation rate (BGR), and random tests.
The BMR is defined as the number of different bits between Alice and Bob divided by the total number of bits of the raw key, calculated using the raw bits before information reconciliation. The BGR is defined as the number of generated raw key bits per packet (bit/pkt). We used the National Institute of Standards and Technology (NIST) test suite~\cite{rukhin2000} to evaluate the randomness of our generated key sequences.

\subsection{Optimal Design Parameters in the CO-SKG Method}
To improve CO-SKG's performance, we conducted extensive experiments to find the optimal parameters in the processes of channel obfuscation and effective key generation.

\begin{figure}
\centering
\subfigure[BMR]{
\includegraphics[scale = 0.24]{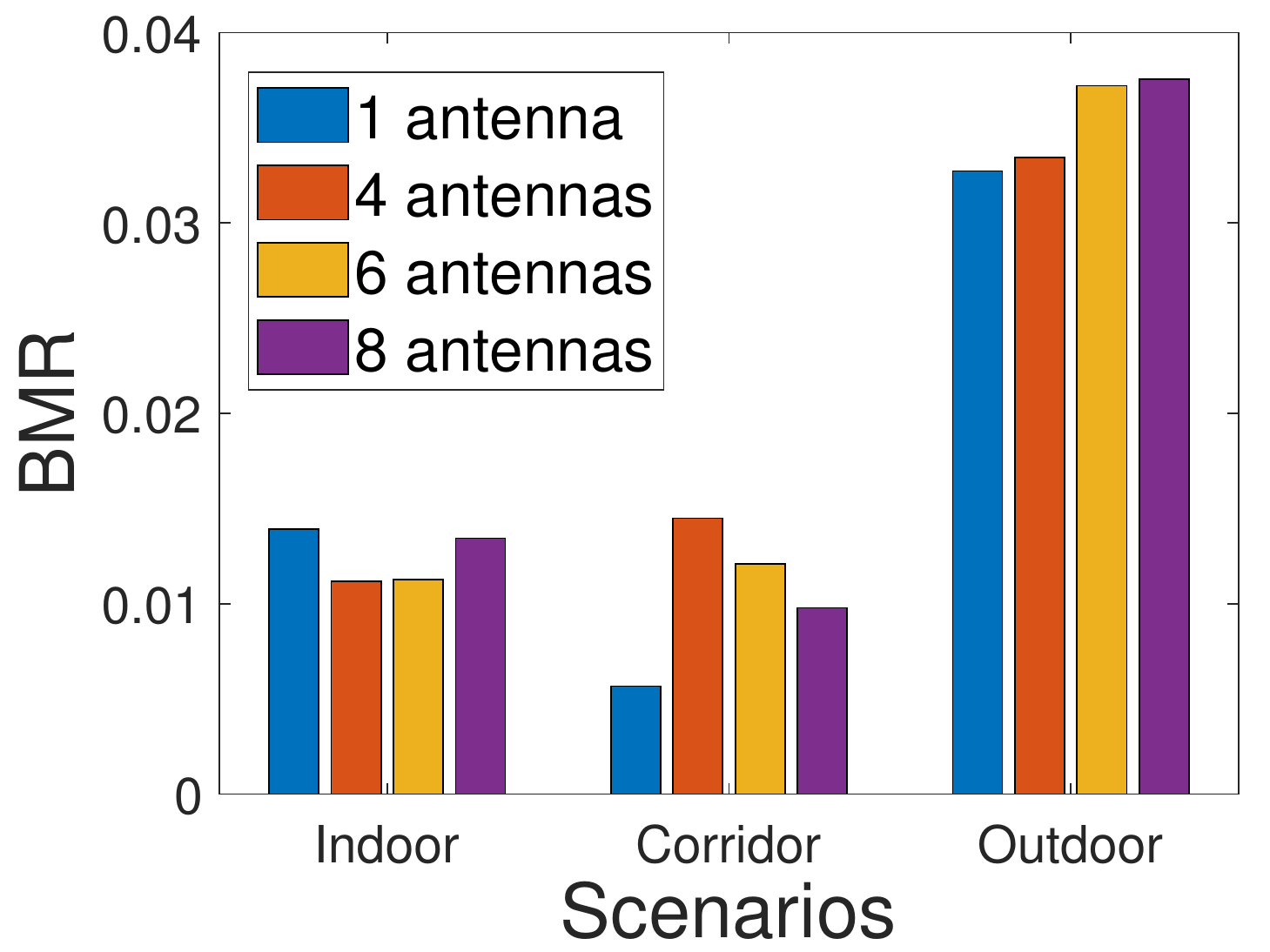}
\label{antbmr}
}
\subfigure[BGR]{
\includegraphics[scale = 0.24]{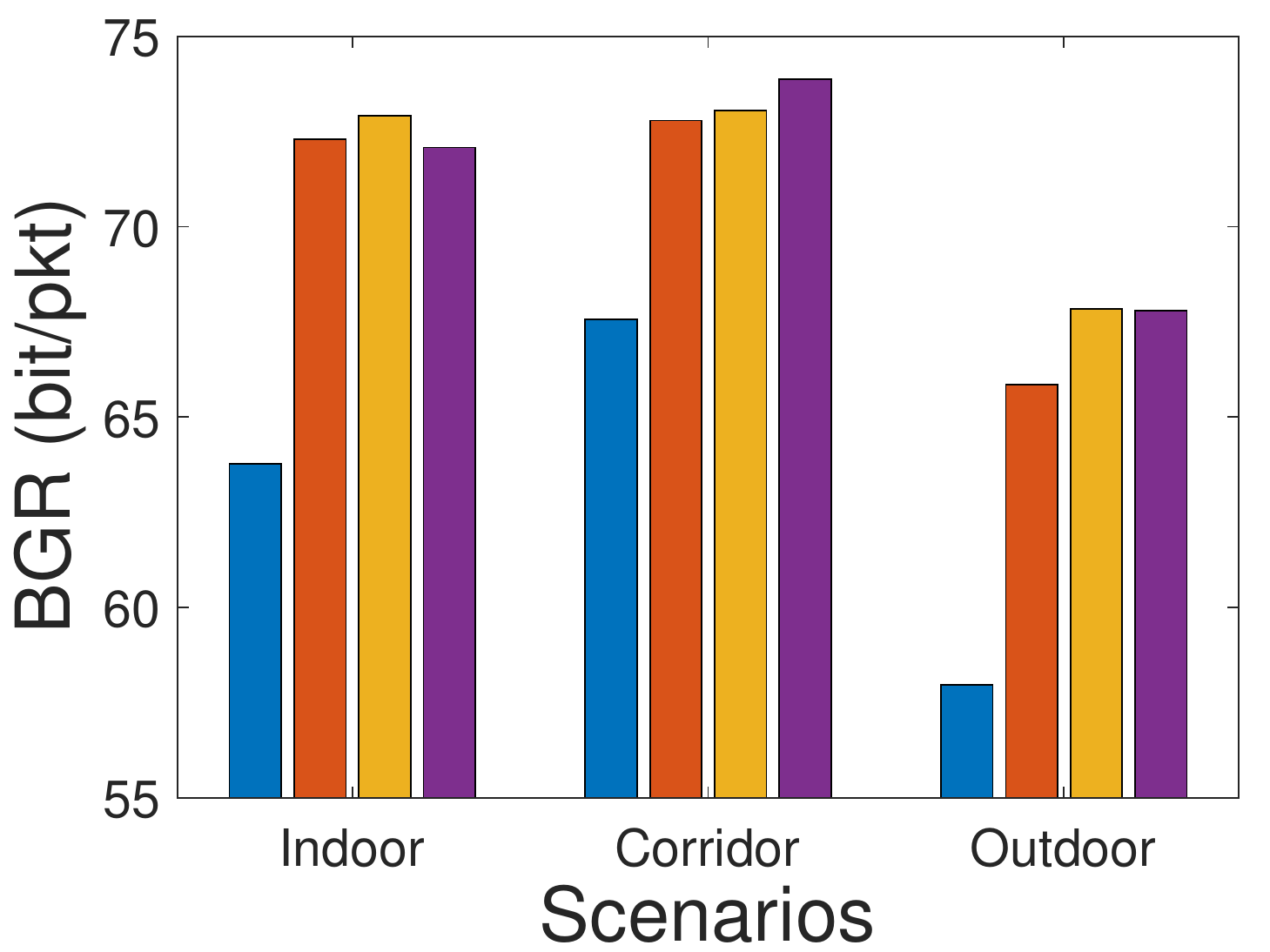}
\label{antbgr}
}
\caption{Performance under different number of antennas.}
\label{ant}
\end{figure}

\subsubsection{Channel Obfuscation} Since random antenna scheduling and random filtering are exploited to realize channel obfuscation. We evaluate the impacts of antenna number and filter length on the performance.
Fig.~\ref{ant} compares the performance of BMR and BGR under different numbers of antennas $M$. 
The impact of $M$ on BMR varies for different scenarios, while for all three scenarios, the BGR grows with the antenna number $M$. In particular, compared with the single antenna case, the BGRs of other antenna numbers have increased by at least $5$ bit/pkt. It is also observed that the BGR levels off when the antenna number becomes large, e.g., $M = 8$. The random filtering has already provided enough randomness, and thus no need to deploy more antennas. According to the result, we choose the antenna number to be $8$ in the design of random antenna scheduling.

\begin{figure}
\centering
\subfigure[BMR]{
\includegraphics[scale = 0.24]{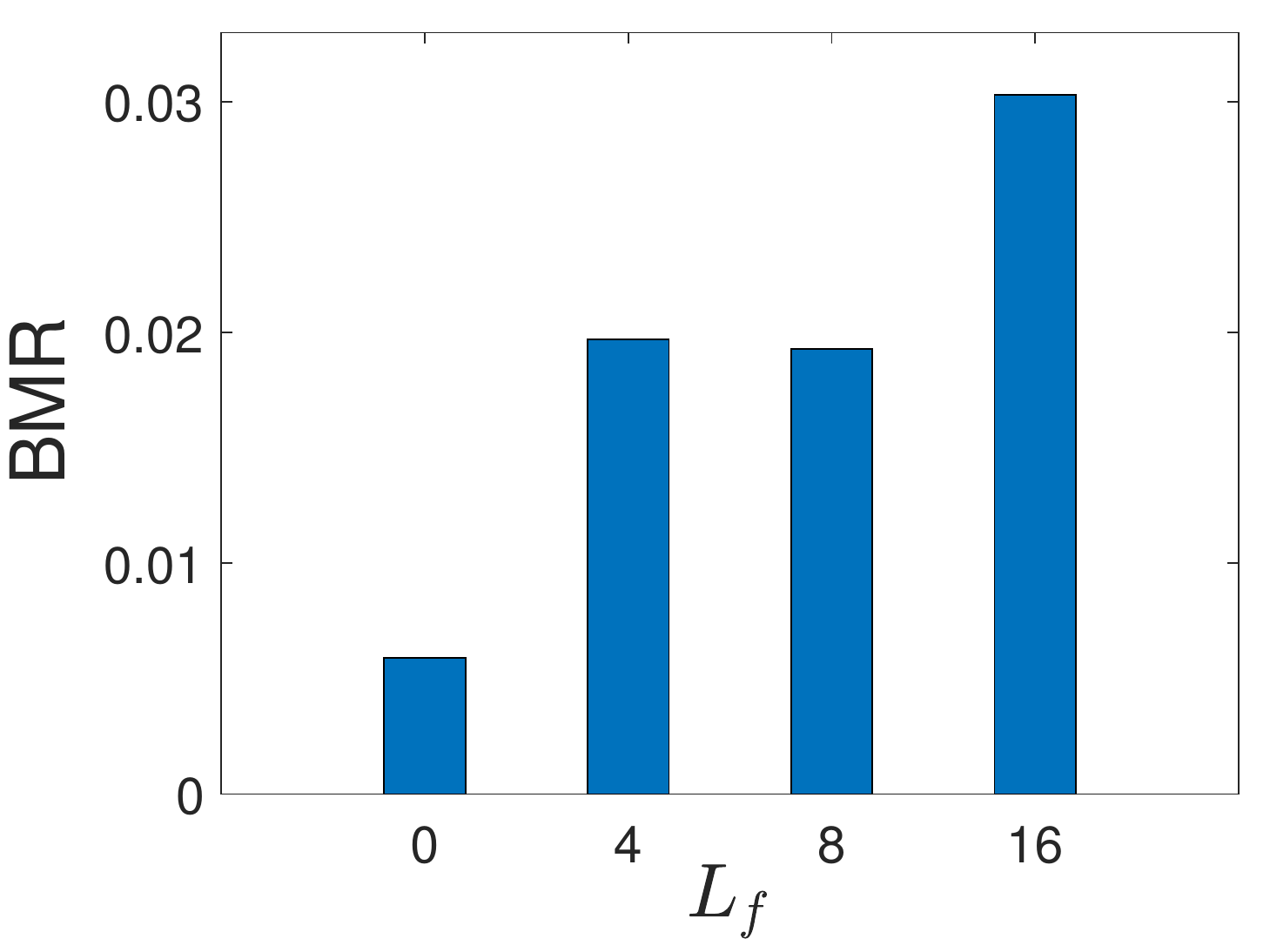}
\label{fliterlenbdr}}
\subfigure[BGR]{
\includegraphics[scale = 0.24]{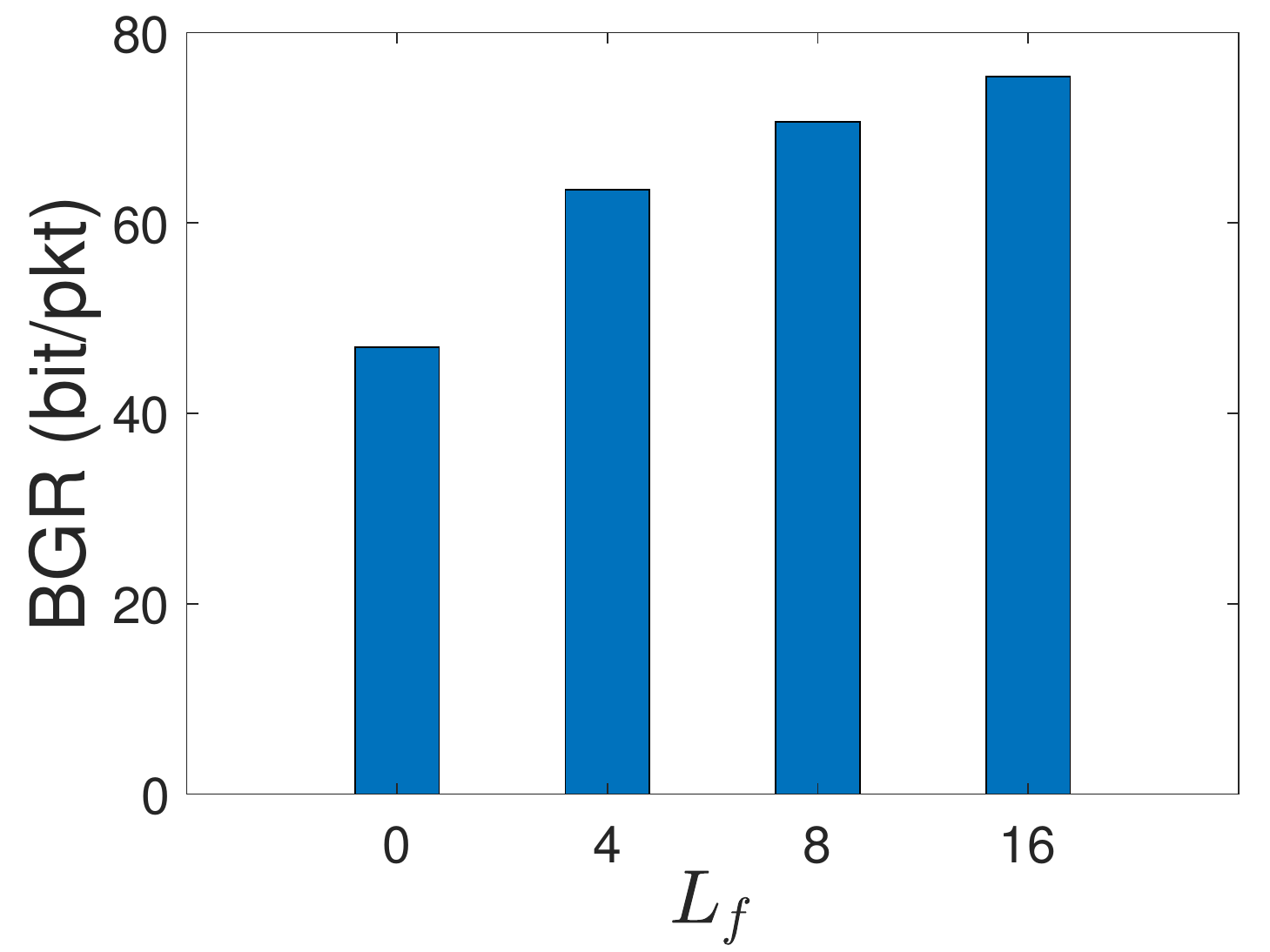}
\label{fliterlenkgr}}
\caption{Performance under different filter length in an indoor scenario.}
\label{fliterlen}
\end{figure}

Fig.~\ref{fliterlen} shows the impact of filter length, $L_f$, on the performance of BMR and BGR. In the experiment, $L_f$ is set as $0$, $4$, $8$ and $16$, where $L_f = 0$ represents the case that random filtering is not used. It is observed that when $L_f$ increases from $0$ to $16$, both BMR and BGR increase. When random filtering is not used, BMR is below $0.01$. When $L_f = 4$, BMR rises rapidly to $0.019$, and then after having a slightly decrease at $L_f = 8$, continuously increases to $0.03$ at$L_f = 16$.
For BGR, when random filtering is not used, BGR is $46.9$ bit/pkt. When $L_f = 4$, BGR also rises rapidly to $63.4$ bit/pkt, and then is slowly increasing to $76.3$ bit/pkt at $L_f = 16$. To make a trade-off between BMR and BGR, we choose the filter length $L_f = 8$ in the design of the random filtering.

\begin{figure}
\centering
\subfigure[BMR]{
\includegraphics[scale = 0.24]{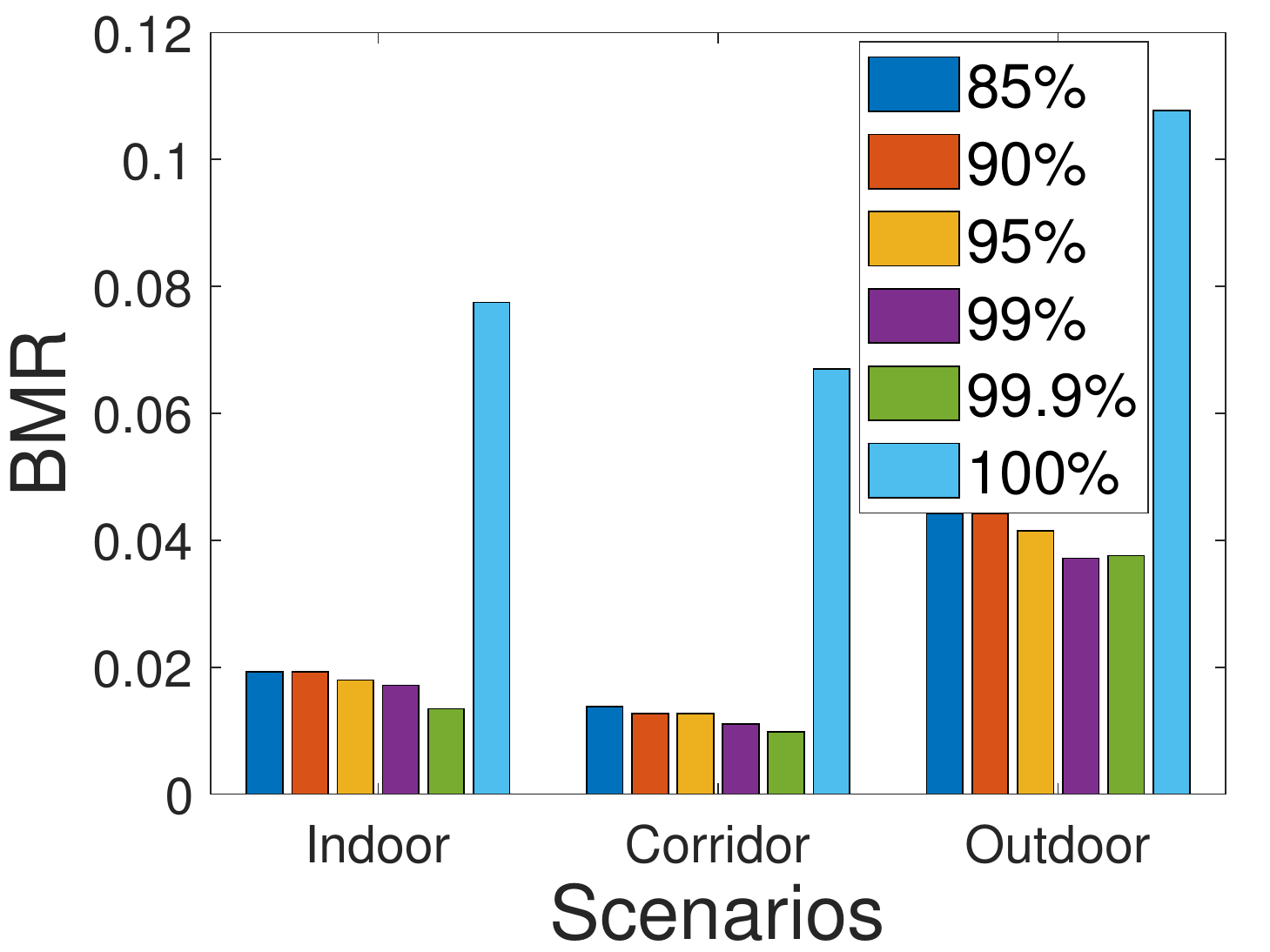}
\label{pcbdr}}
\subfigure[BGR]{
\includegraphics[scale = 0.24]{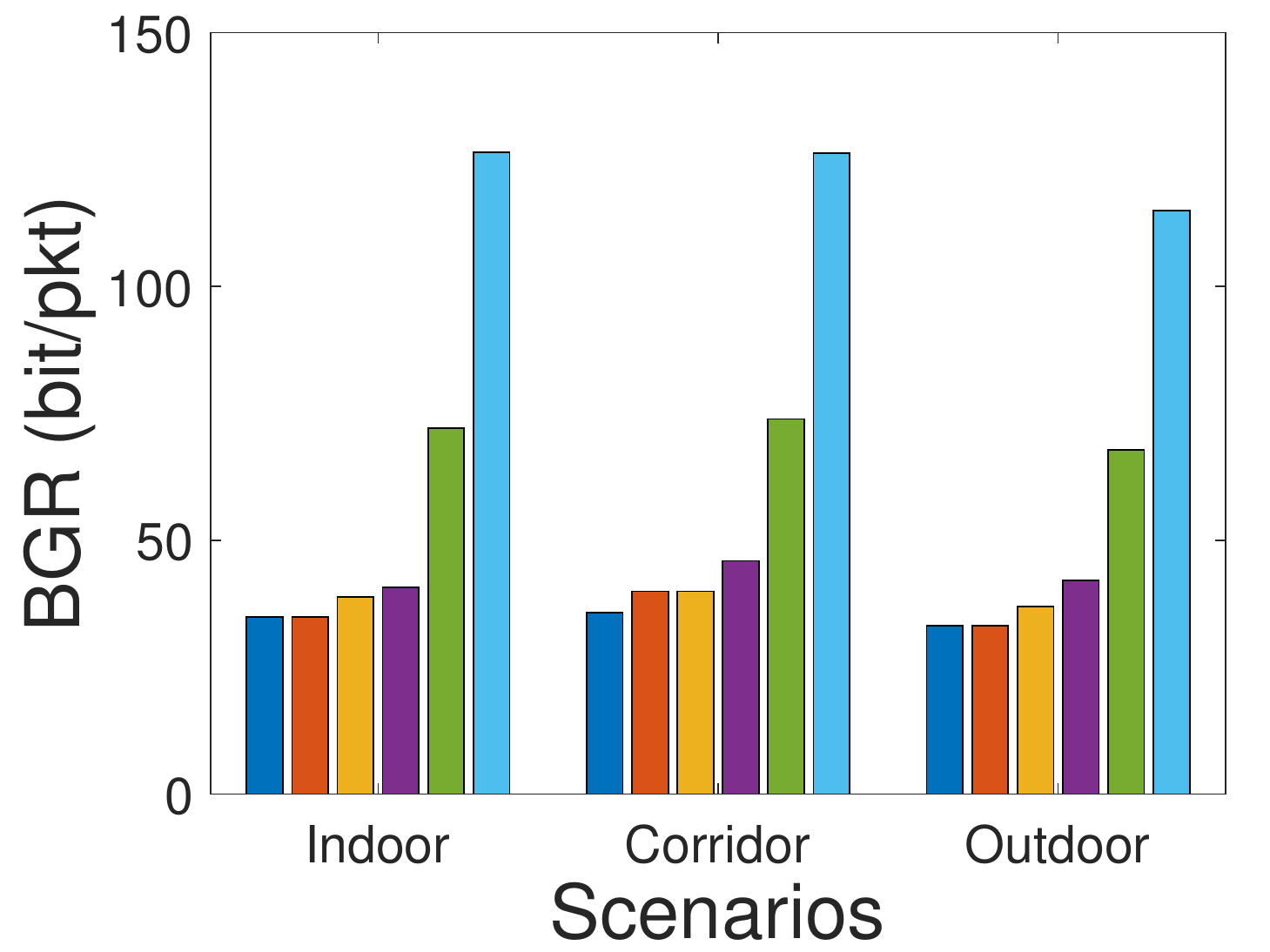}
\label{pckgr}}
\caption{Performance under the impact of proportion of principal components.}
\label{pc}
\end{figure}

\subsubsection{Effective Key Generation} We study the performance of secret key under different design parameters in the K-L transform and adaptive quantization processes. Fig.~\ref{pc} compares the performance of BMR and BGR with different proportions of principal components used for key generation. We specify the percentage of the principal components, $\eta$, to choose the first $P$ principal components. We examine the BMR and BGR by changing $\eta$ from $85\%$ to $100\%$ and find that BGR increases from around $30$ bit/pkt to over $100$ bit/pkt with the increase of $\eta$. The BMR decreases when $\eta$ increases from $85\%$ to $99.9\%$ but increase rapidly over $0.06$ at $100\%$. It is because the rest components are severely corrupted by noise. Therefore, we set the percentage threshold $\eta$ as $99.9\%$ in the K-L transform.

\begin{figure}
\centering
\subfigure[BMR]{
\includegraphics[scale = 0.24]{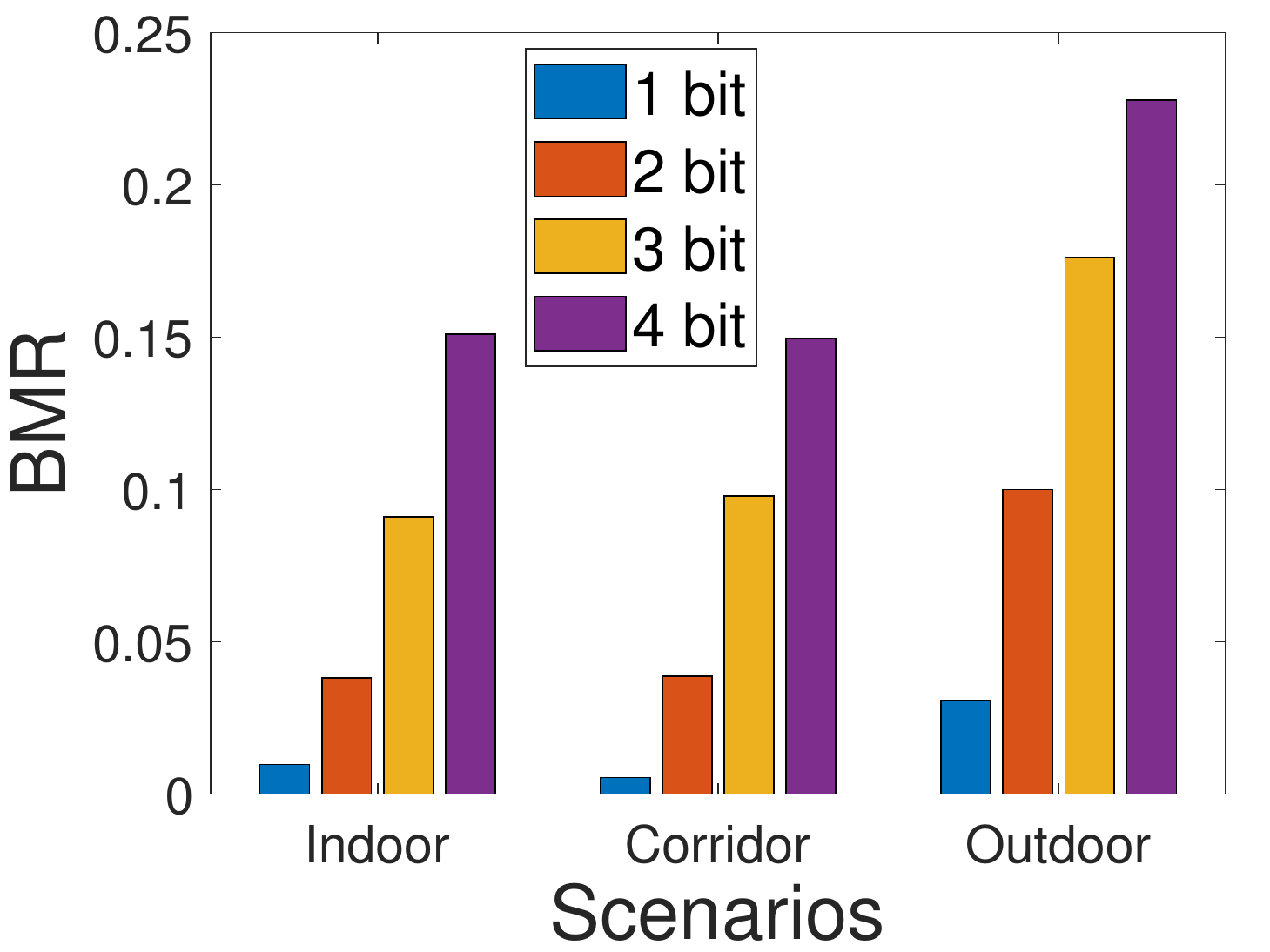}
\label{pv1lvbdr}}
\subfigure[BGR]{
\includegraphics[scale = 0.24]{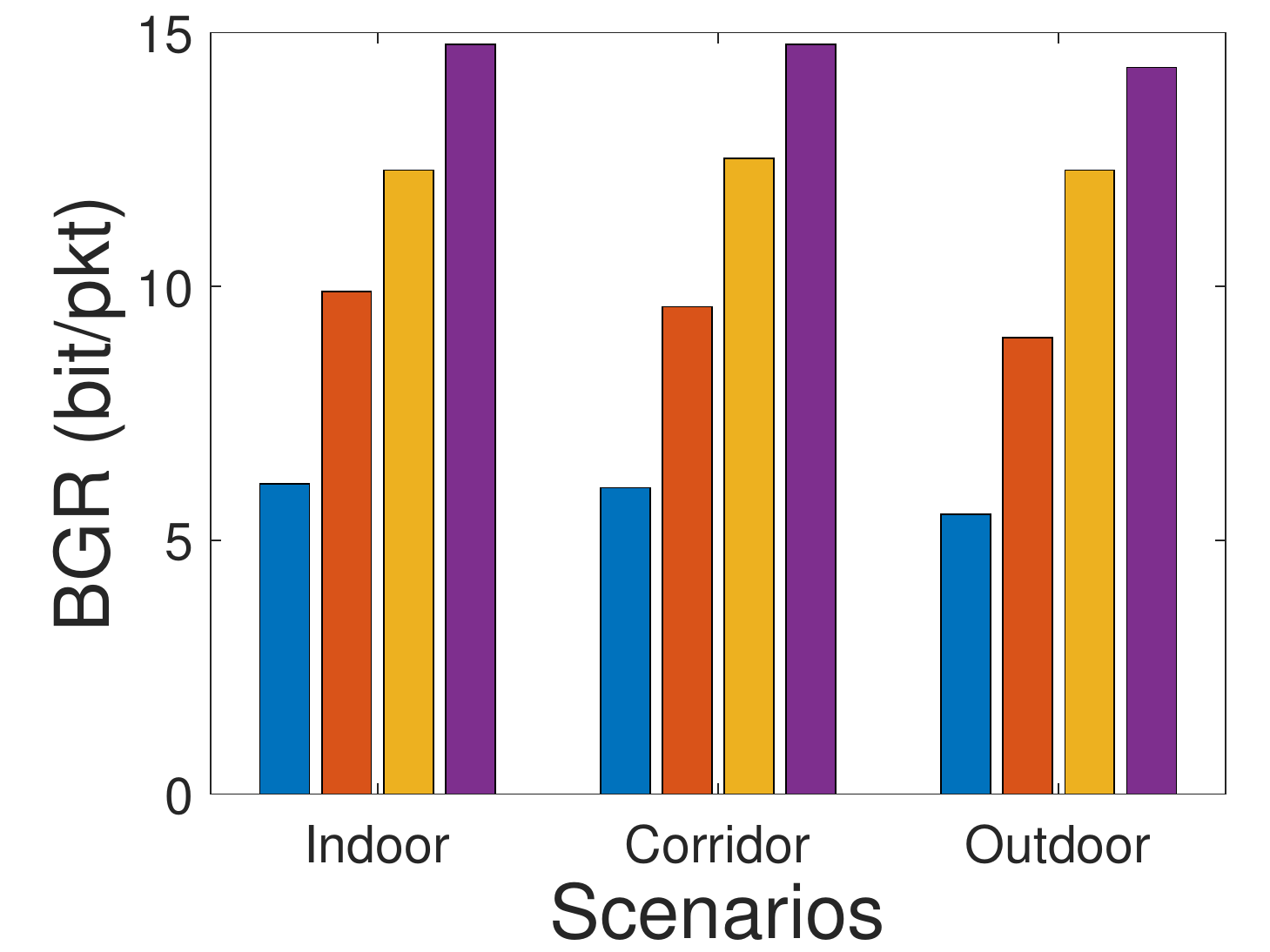}
\label{pc1lvkgr}}
\caption{Performance of the impact of quantization levels of first component.}
\label{pcqlv}
\end{figure}

Next, we perform multi-bit quantization on the first principal component since its energy is much higher than others. We measure the BMR and BGR from $1$ bit to $4$ bits quantization of the first principal component. As shown in Fig.~\ref{pcqlv}, both BGR and BMR grow as the quantization level increases. Since some bits are discarded in the guard band during quantization, the BGR is not a strict double relationship as the quantization level increases. Therefore, we need to balance the impact on BMR and BGR when choosing an appropriate quantization level. According to the experimental results, we set the quantization level for the first principal component as $2$ bits.

\begin{figure}[!t]
\centering
\subfigure[BMR]{
\includegraphics[scale = 0.24]{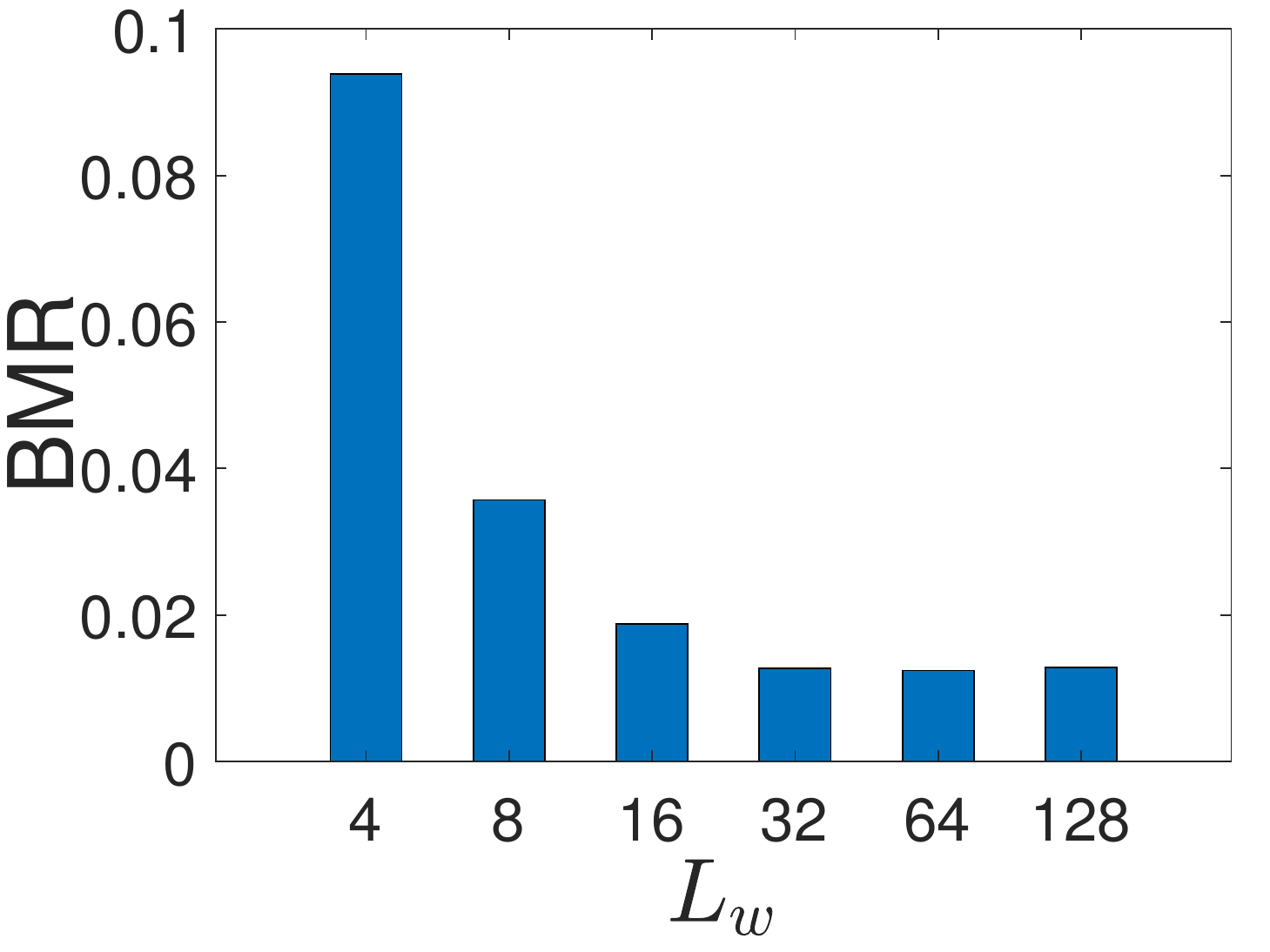}
\label{winbdr}}
\subfigure[BGR]{
\includegraphics[scale = 0.24]{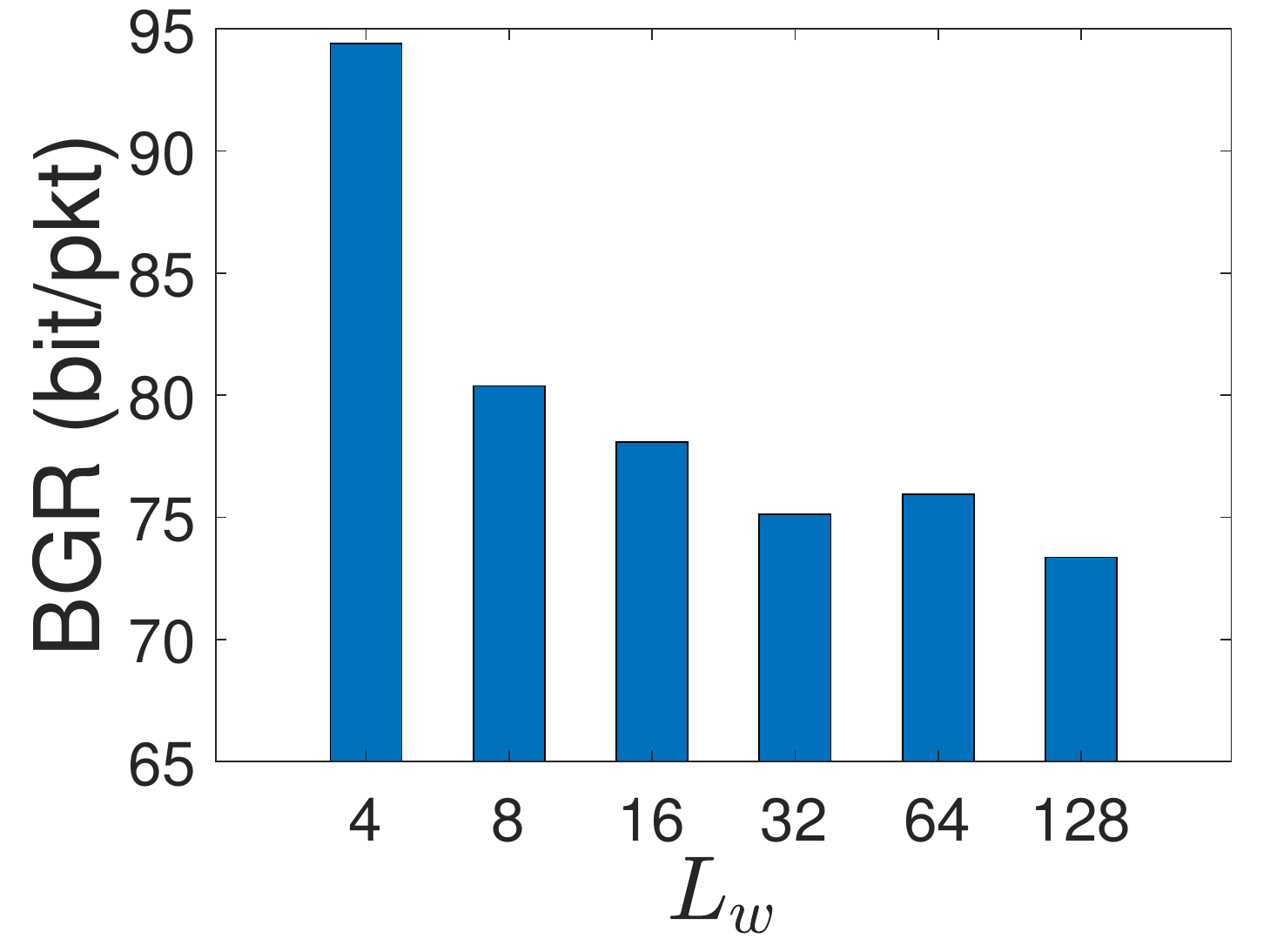}
\label{winkgr}}
\caption{Performance under different window length in an indoor scenario.}
\label{win}
\end{figure}

Finally, Fig. \ref{win} shows the impact of window length, $L_w$, on the key generation performance. We find that BMR decreases from $0.094$ to $0.012$ as $L_w$ increases from $4$ to $64$, and then increases to $0.138$ when $L_w = 128$. BGR decreases from $94.4$ bit/pkt to $71.5$ bit/pkt as $L_w$ increases from $4$ to $128$. To make a trade-off between these metrics, we set the window length $L_w = 64$. Table~\ref{parameters} summarizes above optimal parameters used in the CO-SKG approach.

\begin{table}[!t]
\renewcommand\arraystretch{1.2}
\caption{Optimal parameters of the CO-SKG approach}
\begin{tabular}{l c}
\hline
\hline
Parameters & Values\\
\hline
Number of antennas of Alice, $M$ & $8$ \\
Filter length, $L_f$ & $8$ \\
Energy Percentage of selected principal components, $\eta$ & $99.9\%$ \\
Quantization level of first principal component & $2$ bit \\
Window length, $L_w$ & $64$ \\
\hline
\end{tabular}
\label{parameters}
\end{table}

\begin{figure}[!t]
\centering
\subfigure[BMR]{
\includegraphics[scale = 0.275]{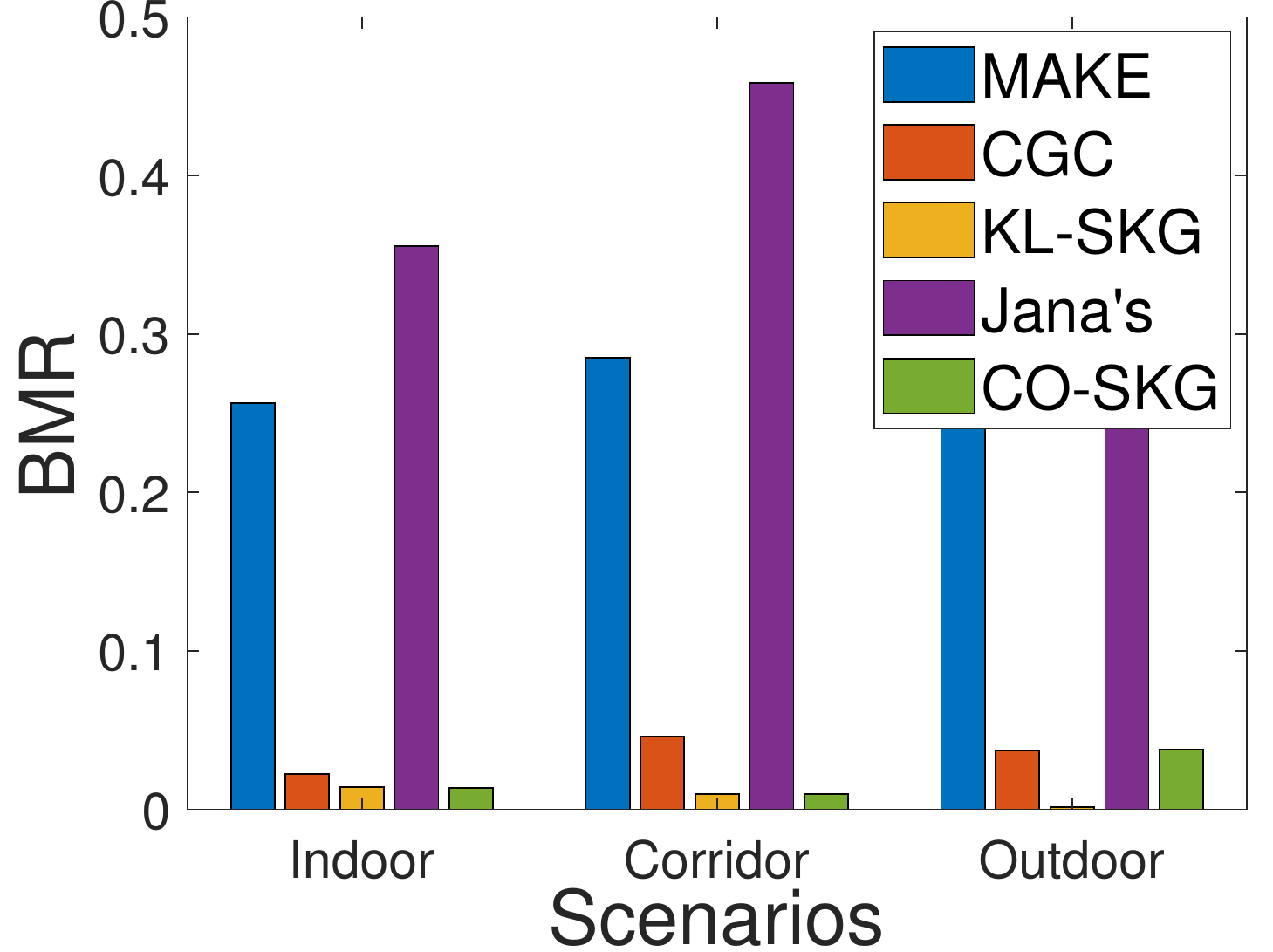}
\label{compappbdr}}
\subfigure[BGR]{
\includegraphics[scale = 0.275]{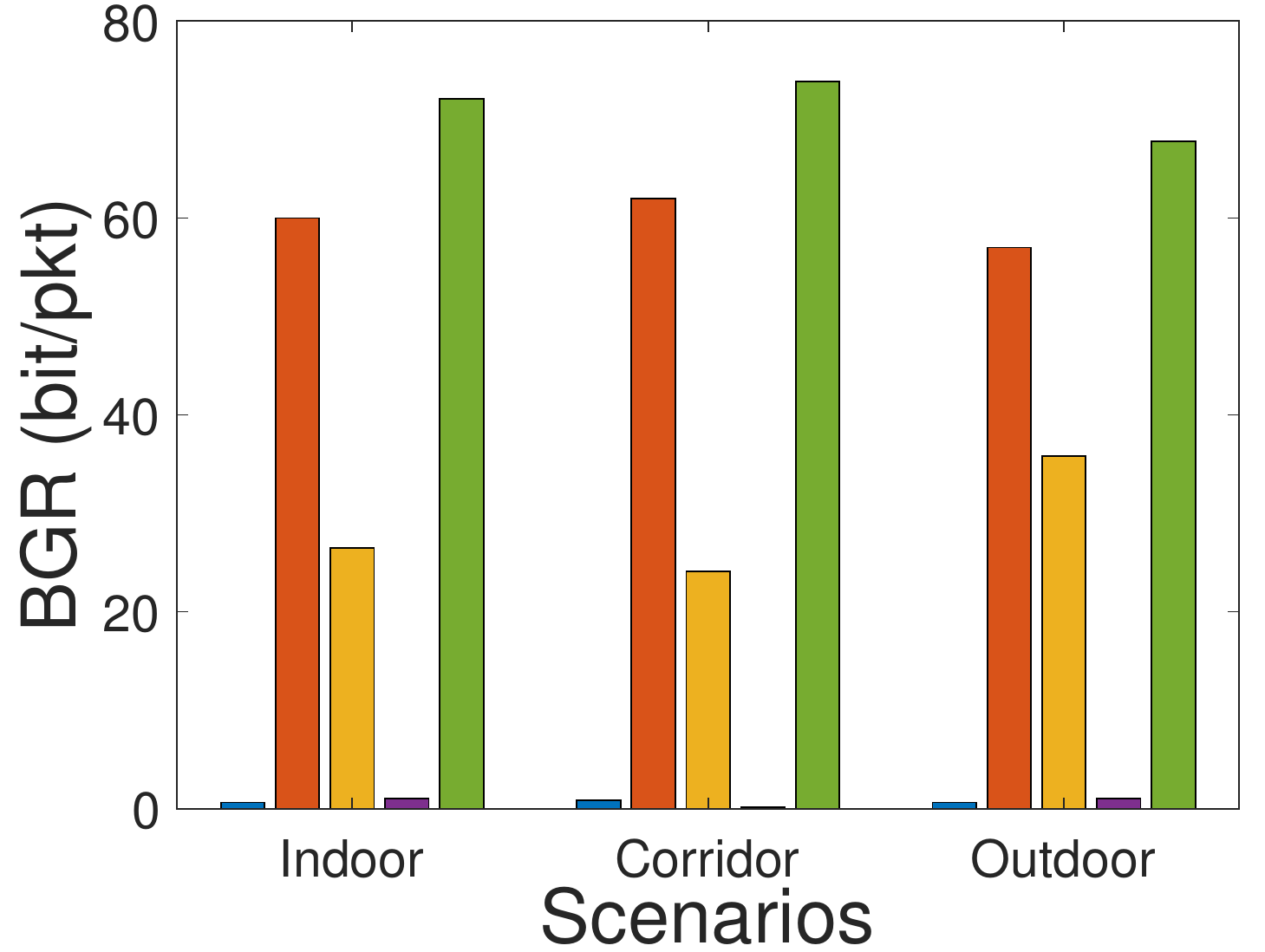}
\label{compappkgr}}
\caption{Performance comparison of different key generation approaches.}
\label{compapp}
\end{figure}

\begin{table}[!t]
\renewcommand\arraystretch{1.2}
\caption{NIST statistical test results}
\begin{tabular}{l c c c c c}
\hline
\hline
Test & CO-SKG & KL-SKG & CGC & Jana's & MAKE\\
\hline
Freq. & 0.4106 & 0.0709 & 1 & 0.7113 & 0.0240\\
Block Freq. & 0.1108 & 0 & 0.0011 & 0.5035 & 0.4371\\
Runs. & 0.4195 & 0 & 0.9500 & 0 & 0.1323\\
Longest of 1's. & 0.1059 & 0 & 0 & 0 & 0.0080\\
\multirow{2}{*}{Serial.} & 0.2118 & 0 & 0 & 0 & 0\\
& 0.3169 & 0 & 0 & 0 & 0.0001\\
Approx. Entropy. & 0.0468 & 0 & 0 & 0 & 0\\
Cumusum & 0.5014 & 0.4433 & 0.4994 & 0.5035 & 0.0204\\
\hline
\end{tabular}
\label{nist_approach}
\end{table}

\subsection{Comparison with Existing Typical Key Generation Methods}
Fig.~\ref{compapp} compares the BMR, and BGR performance of the proposed CO-SKG approach with four existing typical key generation approaches, including MAKE~\cite{MAKE}, CGC~\cite{CGC}, KL-SKG~\cite{PCA} and Jana's~\cite{Jana2009effectiveness}. 
From Fig.~\ref{compappbdr}, MAKE and Jana's have higher BMR, as they use RSSI for key generation, which has slight fluctuation, mainly dominated by the noise in slowly varying environments. CGC, KL-SKG, and CO-SKG extract secret key from CSI, which fluctuates over multiple subcarriers and thus have better performance of the key agreement in the slowly varying environments.
Since KL-SKG and CO-SKG use the K-L transform to reduce noise, they have lower BMRs than CGC. 

Fig.~\ref{compappkgr} shows that CO-SKG provides higher BGR than other approaches, achieving the rate of $72.08$, $73.88$, and $67.79$ bit/pkt for scenarios of the indoor, corridor, and outdoor, respectively.
It is because that CO-SKG exploits both random antenna scheduling and random filtering to increase the CSI fluctuation, so more components can be extracted in K-L transform. Since MAKE and Jana's use RSSI as the channel parameters, they obtain only one sample for each round, so their BGRs are lower than those of CSI-based approaches, i.e., CGC, KL-SKG, and CO-SKG.

In Table~\ref{nist_approach}, we evaluate the randomness of the raw key sequences before privacy amplification using the widely adopted NIST random test suite.
When the P-value test result is greater than the threshold, usually chosen as $0.01$, the sequence passes the
test. The key sequences are extracted in an indoor environment, and its length is set as $1024$ bits. Table~\ref{nist_approach} shows that the proposed CO-SKG approach can pass all of these seven tests, while other approaches can only pass part of them.

The above experimental results have verified that the proposed CO-SKG can provide secret keys with higher key agreement, faster key generation rate, and more substantial randomness in slowly varying environments than existing approaches.

\section{Security Evaluation}
\label{sec:SA}
In this section, we evaluate the security of the proposed CO-SKG approach against the four attacks described in Sec.~\ref{sec:attack}.
\subsection{Preventing Predictable Channel Attack}
CO-SKG prevents Eve from causing the predictable changes in the channel measurements by controlling the movements of some intermediate objects. 
According to (\ref{eq:gkn}), the variation of $\hat{H}_u(k,n)$ depends on the physical channel, as well as the selected antenna and the filter coefficients.
Therefore, the pattern of variation of $\hat{H}_u(k,n)$ does not follow the movements of Eve, even in a slowly varying environment.
\begin{figure}
\centering
\subfigure[Original]{
\includegraphics[scale = 0.275]{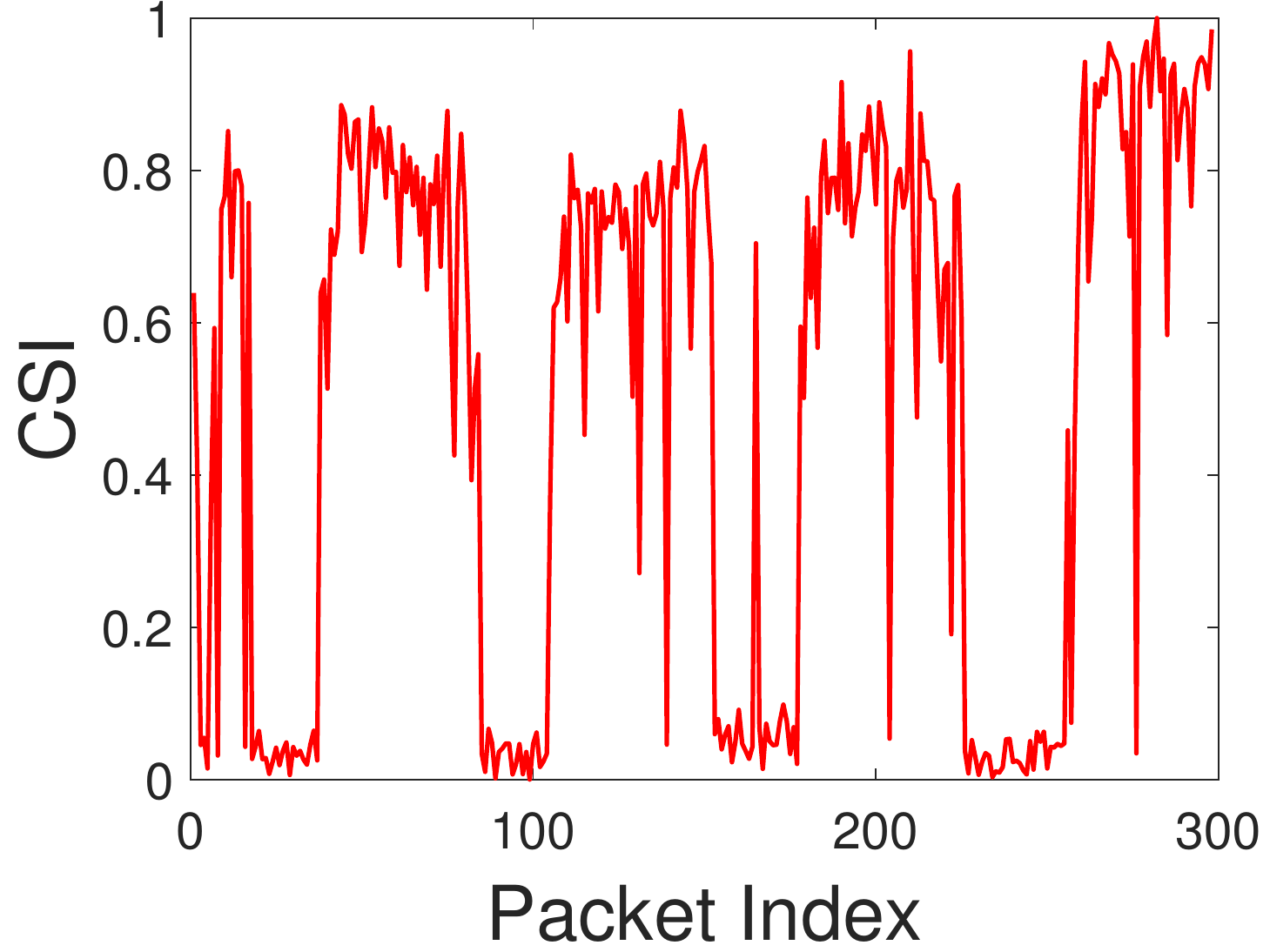}
\label{preattack1}}
\subfigure[CO-SKG]{
\includegraphics[scale = 0.275]{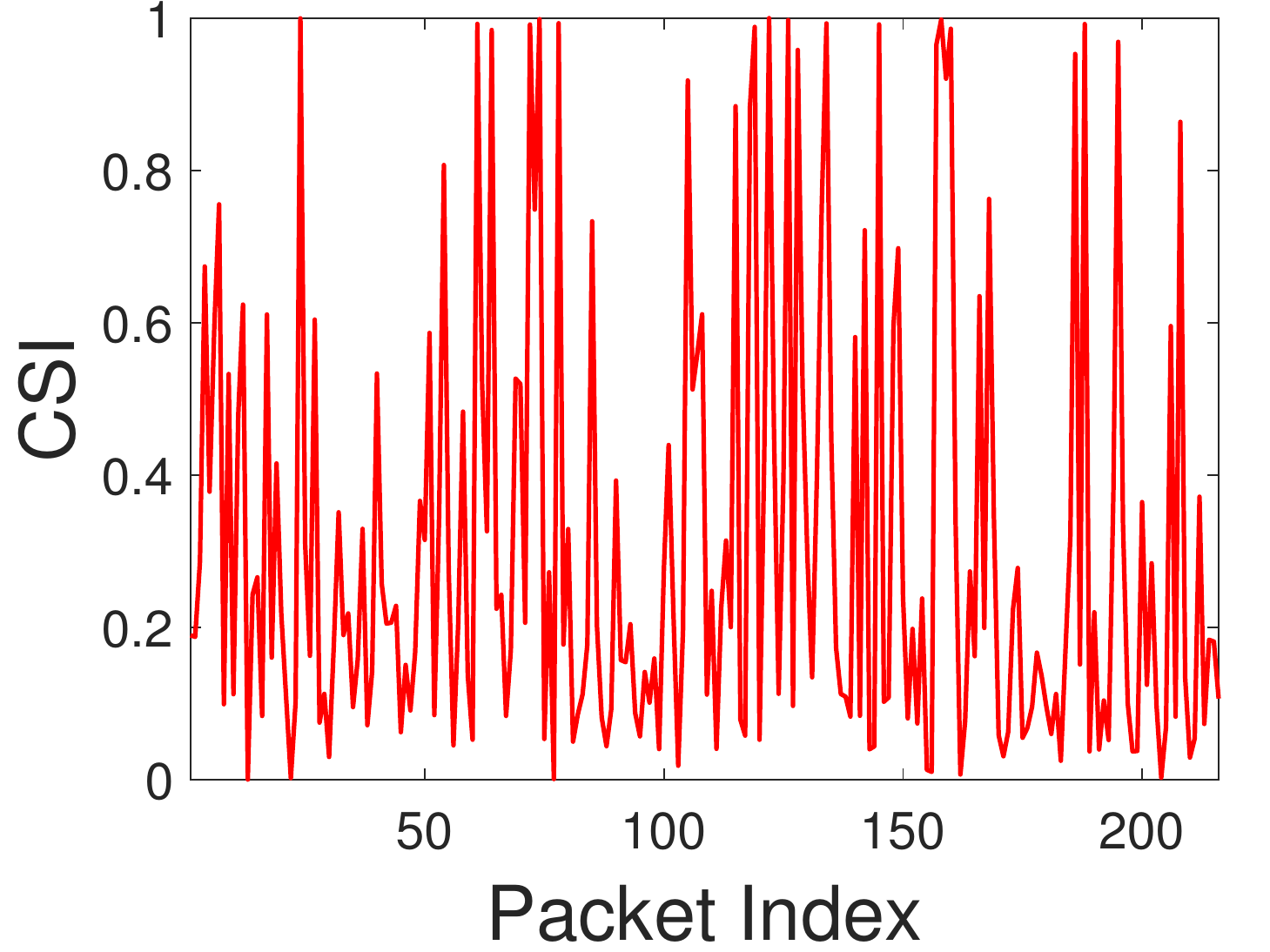}
\label{preattack2}}
\caption{The amplitude of CSI under predictable channel attack in an indoor scenario.}
\label{preattack}
\end{figure}

We evaluate the performance in a corridor environment, where Alice and Bob are placed on the two sides of the corridor with a door in the middle. The predictable channel attack is implemented by opening and closing the door periodically over $2$ minutes.
Fig.~\ref{preattack} compares the amplitude of CSI sequences on the $256$-th subcarrier of CO-SKG against the original one without channel obfuscation. The pattern of variation in Fig.~\ref{preattack1} is highly related to the opening or closing state of the door, while the relevance is primarily reduced in Fig.~\ref{preattack2}. In particular, when the door is closed, the channel becomes non-line-of-sight (NLoS), and thus the CSI amplitude in Fig.~\ref{preattack1} is significantly less than that in the line-of-sight (LoS) case in which the door is opening. The CSI amplitude in Fig.~\ref{preattack2} has a significant fluctuation under both cases, which indicates that CO-SKG can resist the predictable channel attack.

%

\subsection{Preventing Position Replay Attack}
Since $\hat{H}_u(k,n)$ is not determined solely upon the locations of Alice and Bob, CO-SKG can prevent Eve from obtaining the same key via location replay attacks. 
When Eve moves to the same position at another time round of $k'$, it will obtain $\hat{H}_E(k',n)$, which is independent from $\hat{H}_B(k,n)$, as $\alpha_{k,n}$ and $m'$ changes independently over $k$.
\begin{figure}
\subfigure[$\rho_{BE}$]{
\includegraphics[scale = 0.275]{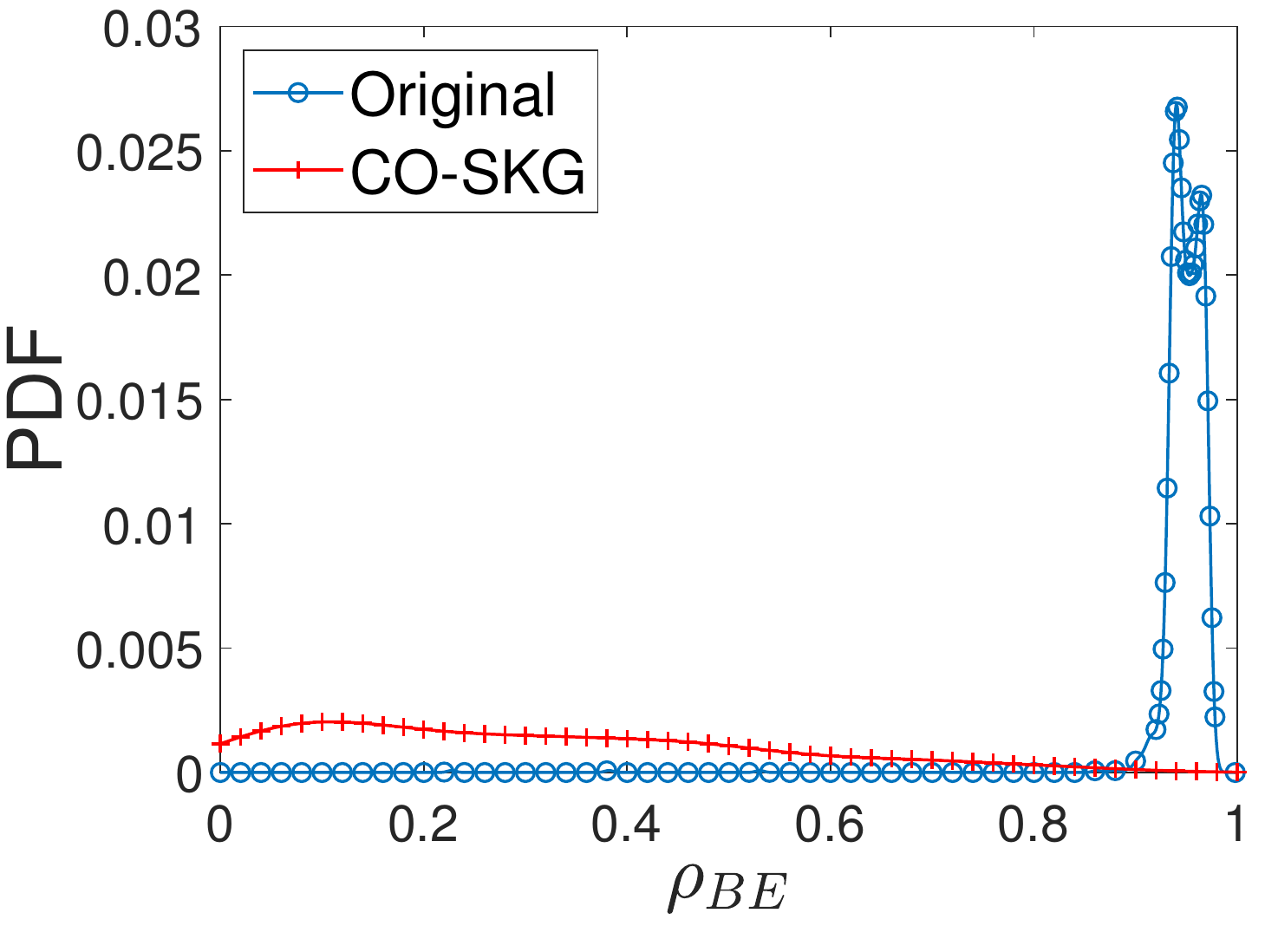}
\label{replay}
}
\subfigure[$N_d$]{
{\includegraphics[scale = 0.275]{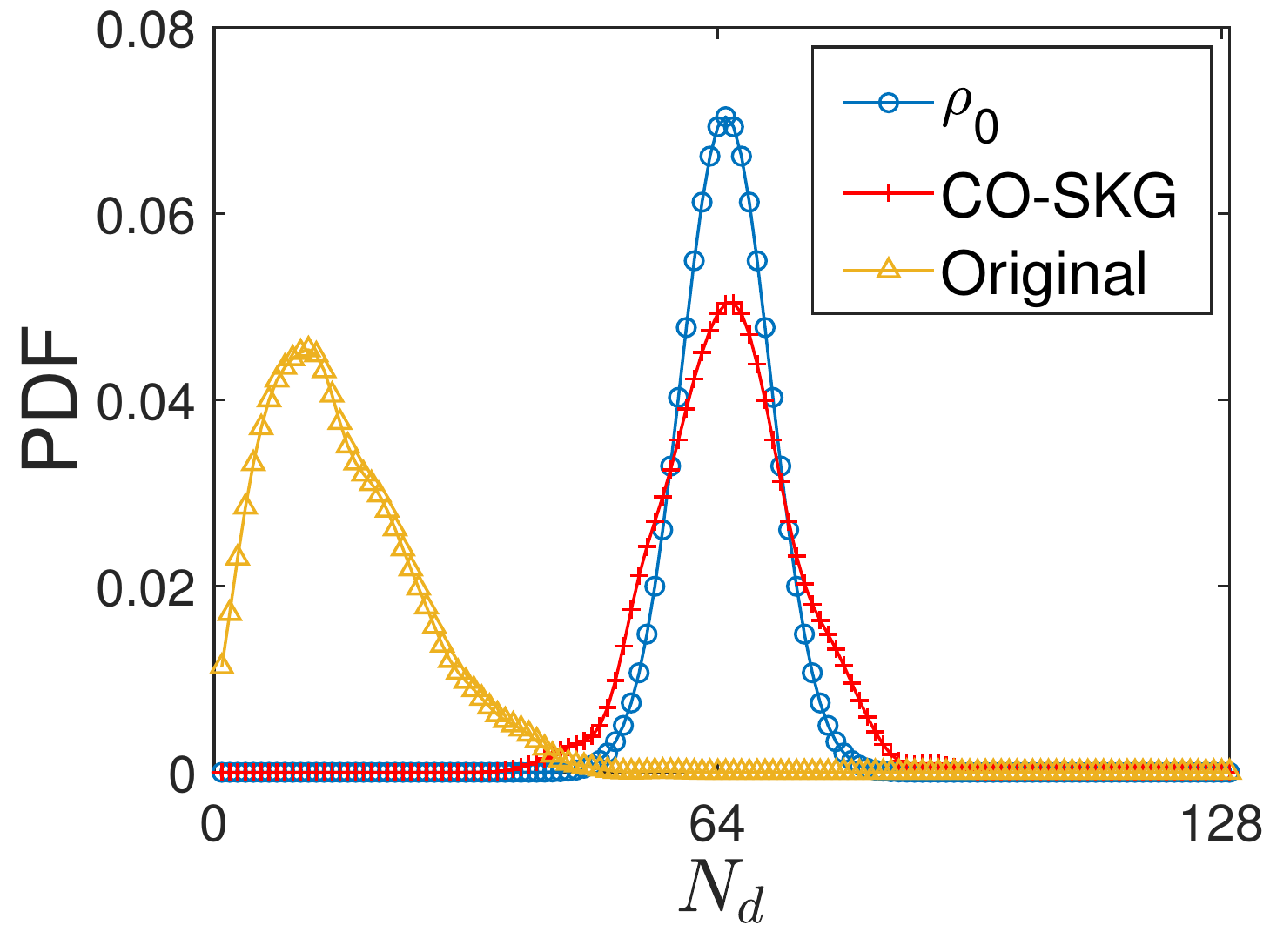}}
\label{fig:kcr}
}
\caption{The probability distribution of $\rho_{BE}$ and $N_d$ in an indoor scenario.}
\end{figure}

We implemented the position replay attacks in an indoor scenario. After Alice and Bob communicate for a while, Bob moves to a new location, and Eve quickly moves to Bob's original position and attempts to obtain a similar CSI as that of Bob.
Fig.~\ref{replay} reports the probability distribution of the correlation coefficient over all subcarriers, $\rho_{BE}$.
%
It is observed that the value of $\rho_{BE}$ is concentrated at the range from $0.9$ to $1$ for the original approach, while its value is largely reduced to below $0.5$ for the CO-SKG approach, which indicates that CO-SKG can effectively defend against the position replay attack in a slowly varying environment.

\subsection{Preventing Effective Brute-Force Attack}
CO-SKG prevents the Eve from shortening the time-complexity through the effective brute-force attack, which is described in Section~\ref{sec:attack}.
By exploiting channel obfuscation, CSI samples will be changed irregularly and thus reduce the proportion of repeated bit segments in ${\bf q}_B$. We divide ${\bf q}_B$ into multiple groups, each of which contains $128$ bits.
The number of different bits between adjacent groups, $N_d$, is defined as $N_d = ||{\bf q}_B^{j+1} - {\bf q}_B^{j}||_1$, where $j$ is the index of group and $||\cdot||_1$ represents the norm-1 function. Theoretically, when the raw keys in adjacent groups are independent, the probability distribution function of $N_d$ should be $\rho_0 =C_{128}^{N_d}\left(\frac{1}{2}\right)^{128}$.
Fig.~\ref{fig:kcr} compares the probability distribution of $N_d$ that is calculated from the raw keys generated by the original approach, the CO-SKG approach, and the theoretical function of $\rho_0$. It is observed that the value of $N_d$ in the original approach is concentrated around $18$, which deviates largely from the theoretical value $\rho_0$. In this case, Eve is able to infer the raw key in adjacent groups by the effective brute-force attack. Conversely, the probability distribution of $N_d$ for the CO-SKG approach is close to that of the $\rho_0$, which indicates the effectiveness of CO-SKG in reducing the proportion of repeated bit segments in raw keys. In this way, CO-SKG can resist an effective brute-force attack.

\subsection{Preventing Order Speculation Attack}\label{sec:pos}
\begin{figure}
\centering
\includegraphics[scale = 0.335]{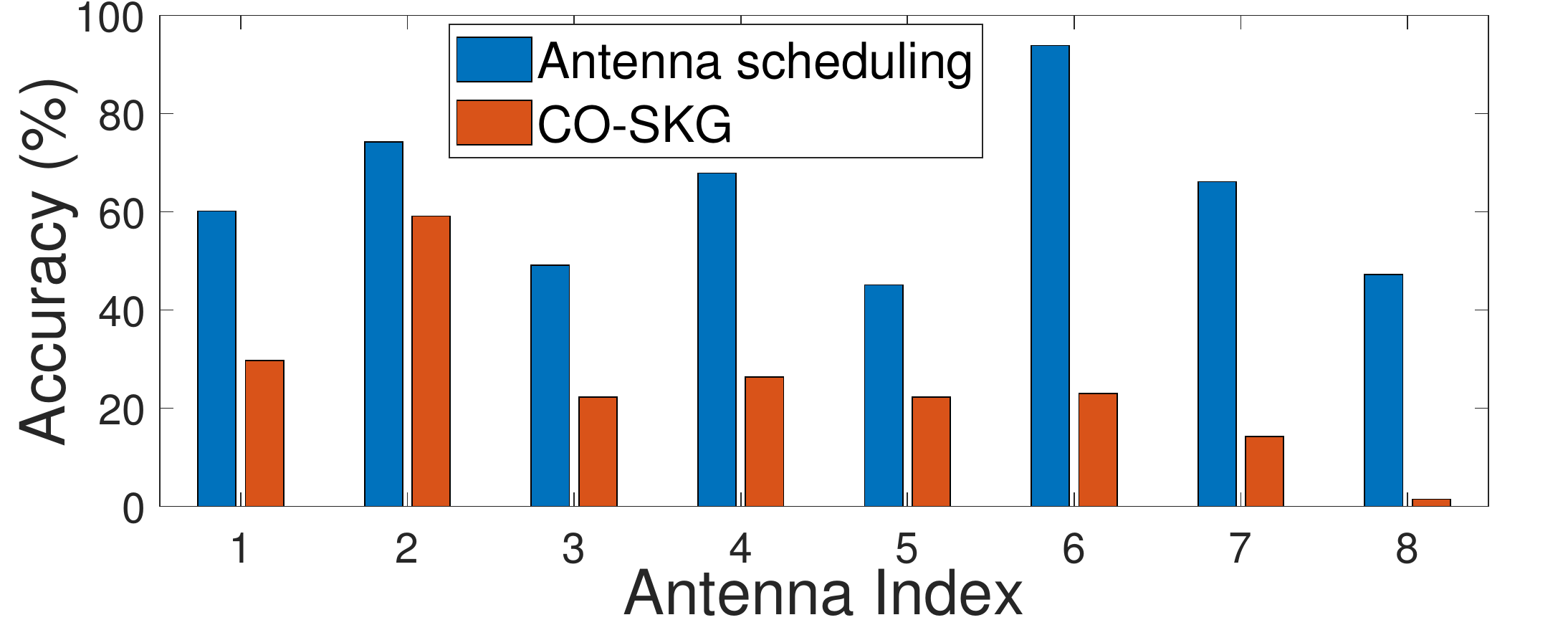}
\caption{Speculation accuracy of Eve in an indoor scenario.}
\label{eveas}
\end{figure}

By exploiting two degrees of freedom, i.e., $m_k$ and $\alpha(k,n)$, CO-SKG prevents Eve from speculating the obfuscation information.
We evaluate the resistance of CO-SKG under this attack in comparison with the approach solely using random $m_k$, which is referred to as the antenna scheduling approach.
Eve implemented the order speculation attack with $1000$ rounds of experimental data collected in an indoor environment. In each round, Eve obtained a CSI vector with $512$ CSI values from OFDM subcarriers.
Eve intends to speculate the index of the used antenna in the current time around by matching its CSI vector with those obtained from previous rounds. The K-means algorithm~\cite{kmeans} is exploited by Eve to partition its unlabelled CSI vectors into eight distinct groupings, corresponding to $M=8$ antennas.
The speculation is defined as accurate when the output index of the K-means algorithm is equal to the index of antennas in use. Fig.~\ref{eveas} compares the speculation accuracy of Eve on different antennas, and the results show that Eve can speculate the antenna index with high accuracy in the antenna scheduling approach, while the accuracy is largely reduced in the proposed CO-SKG approach.

\section{Conclusion}
In this paper, we proposed a key generation protocol, CO-SKG, that incorporates channel obfuscation and effective key generation to provide fast and secure key generation in slowly varying environments.
CO-SKG exploits two degrees of freedom, i.e., antenna index and filter coefficients, to obfuscate channel parameters and hide the obfuscation information. In addition, the joint design of K-L transform and adaptive quantization improves key agreement and randomness effectively.
We implemented CO-SKG using USRP SDR platforms and realizing antenna scheduling with an SP8T switch.
Extensive experiments were conducted, and the results demonstrate that compared with existing typical approaches, CO-SKG can provide higher key agreement, faster key generation rate, and more substantial randomness in three slowly varying scenarios.
Finally, experimental results have verified that our protocol achieves a high-security level against various attacks, including the predictable channel attack and position replay attack from active attackers and effective brute-force attack and order speculation attack of passive attackers.

\bibliographystyle{IEEEtran}
\bibliography{IEEEabrv,Ref}

\begin{thebibliography}{10}
\providecommand{\url}[1]{#1}
\csname url@samestyle\endcsname
\providecommand{\newblock}{\relax}
\providecommand{\bibinfo}[2]{#2}
\providecommand{\BIBentrySTDinterwordspacing}{\spaceskip=0pt\relax}
\providecommand{\BIBentryALTinterwordstretchfactor}{4}
\providecommand{\BIBentryALTinterwordspacing}{\spaceskip=\fontdimen2\font plus
\BIBentryALTinterwordstretchfactor\fontdimen3\font minus
  \fontdimen4\font\relax}
\providecommand{\BIBforeignlanguage}[2]{{%
\expandafter\ifx\csname l@#1\endcsname\relax
\typeout{** WARNING: IEEEtran.bst: No hyphenation pattern has been}%
\typeout{** loaded for the language `#1'. Using the pattern for}%
\typeout{** the default language instead.}%
\else
\language=\csname l@#1\endcsname
\fi
#2}}
\providecommand{\BIBdecl}{\relax}
\BIBdecl

\bibitem{ZHANG2020Frontier}
J.~Zhang, G.~Li, A.~Marshall, A.~Hu, and L.~Hanzo, ``A new frontier for {IoT}
  security emerging from three decades of key generation relying on wireless
  channels,'' \emph{IEEE Access}, vol.~8, pp. 138\,406--138\,446, 2020.

\bibitem{li2019physical}
G.~Li, C.~Sun, J.~Zhang, E.~Jorswieck, B.~Xiao, and A.~Hu, ``Physical layer key
  generation in {5G} and beyond wireless communications: Challenges and
  opportunities,'' \emph{Entropy}, vol.~21, p. 497, 2019.

\bibitem{9328938}
Suwadi, Wirawan, and M.~Yuliana, ``Performance evaluation of secret key
  generation system for static and dynamic condition,'' in \emph{IEEE ICC},
  Batam, Indonesia, Dec. 2020, pp. 423--428.

\bibitem{Jana}
S.~N. Premnath, S.~Jana, J.~Croft, P.~L. Gowda, M.~Clark, S.~K. Kasera,
  N.~Patwari, and S.~V. Krishnamurthy, ``Secret key extraction from wireless
  signal strength in real environments,'' \emph{{IEEE} Trans. Mobile Comput.},
  vol.~12, no.~5, pp. 917--930, 2013.

\bibitem{mathur08}
S.~Mathur, W.~Trappe, N.~Mandayam, C.~Ye, and A.~Reznik, ``Radio-telepathy:
  extracting a secret key from an unauthenticated wireless channel,'' in
  \emph{Proc. ACM MobiCom}, San Francisco, California, USA, Sep. 2008, pp.
  128--139.

\bibitem{KEEPJ}
W.~Xi, M.~Duan, X.~Bai, K.~Zhao, L.~Mo, and J.~Zhao, ``Keep: Secure and
  efficient communication for distributed iot devices,'' \emph{{IEEE} Internet
  Things J.}, pp. 1--1, 2020.

\bibitem{2016Instant}
W.~Xi, C.~Qian, J.~Han, K.~Zhao, S.~Zhong, X.-Y. Li, and J.~Zhao, ``Instant and
  robust authentication and key agreement among mobile devices,'' in
  \emph{Proc. ACM CCS}, New York, NY, USA, Oct. 2016, pp. 616--627.

\bibitem{Bipartite}
H.~Liu, Y.~Wang, Y.~Ren, and Y.~Chen, ``Bipartite graph matching based secret
  key generation,'' in \emph{Proc. IEEE INFOCOM}, Virtual Conference, May.
  2021, pp. 1--9.

\bibitem{Aldaghri_2020}
N.~Aldaghri and H.~Mahdavifar, ``Physical layer secret key generation in static
  environments,'' \emph{{IEEE} Trans. Inf. Forensics Security}, vol.~15, pp.
  2692--2705, Feb. 2020.

\bibitem{6716049}
H.~Zhou, L.~M. Huie, and L.~Lai, ``Secret key generation in the two-way relay
  channel with active attackers,'' \emph{{IEEE} Trans. Inf. Forensics
  Security}, vol.~9, no.~3, pp. 476--488, 2014.

\bibitem{Low-entropy}
P.~Staat, H.~Elders-Boll, M.~Heinrichs, R.~Kronberger, C.~Zenger, and C.~Paar,
  ``Intelligent reflecting surface-assisted wireless key generation for
  low-entropy environments,'' \emph{ArXiv}, vol. abs/2010.06613, 2020.

\bibitem{2021Sum}
G.~Li, C.~Sun, E.~A. Jorswieck, and et~al., ``Sum secret key rate maximization
  for {TDD} multi-user massive {MIMO} wireless networks,'' \emph{{IEEE} Trans.
  Inf. Forensics Security}, vol.~16, pp. 968--982, 2021.

\bibitem{MAKE}
K.~Zeng, D.~Wu, A.~Chan, and P.~Mohapatra, ``Exploiting multiple-antenna
  diversity for shared secret key generation in wireless networks,'' in
  \emph{Proc. IEEE INFOCOM}, San Diego, CA, USA, Mar. 2010, pp. 1--9.

\bibitem{QinExploiting}
D.~Qin and D.~Zhi, ``Exploiting multi-antenna non-reciprocal channels for
  shared secret key generation,'' \emph{{IEEE} Trans. Inf. Forensics Security},
  vol.~11, no.~12, pp. 2693--2705, Dec. 2016.

\bibitem{Applying_12TIFS}
M.~G. Madiseh, S.~W. Neville, and M.~L. McGuire, ``Applying beamforming to
  address temporal correlation in wireless channel characterization-based
  secret key generation,'' \emph{{IEEE} Trans. Inf. Forensics Security},
  vol.~7, no.~4, pp. 1278--1287, 2012.

\bibitem{5934929}
Q.~Wang, H.~Su, K.~Ren, and K.~Kim, ``Fast and scalable secret key generation
  exploiting channel phase randomness in wireless networks,'' in \emph{Proc.
  IEEE INFOCOM}, Shanghai, China, Apr. 2011, pp. 1422--1430.

\bibitem{CGC}
H.~Liu, Y.~Wang, J.~Yang, and Y.~Chen, ``Fast and practical secret key
  extraction by exploiting channel response,'' in \emph{Proc. IEEE INFOCOM},
  Turin, Italy, Apr. 2013, pp. 3048--3056.

\bibitem{6567033}
P.~Huang and X.~Wang, ``Fast secret key generation in static wireless networks:
  A virtual channel approach,'' in \emph{Proc. IEEE INFOCOM}, Turin, Italy,
  Apr. 2013, pp. 2292--2300.

\bibitem{8543098}
Y.~Huang, L.~Jin, H.~Wei, Z.~Zhong, and S.~Zhang, ``Fast secret key generation
  based on dynamic private pilot from static wireless channels,'' \emph{China
  Communications}, vol.~15, no.~11, pp. 171--183, 2018.

\bibitem{liguyue2017}
G.~Li, A.~Hu, J.~Zhang, and B.~Xiao, ``Security analysis of a novel artificial
  randomness approach for fast key generation,'' in \emph{Proc. IEEE GLOBECOM},
  Singapore, Dec. 2017, pp. 1--6.

\bibitem{Gui2015Untrusted}
R.~Guillaume, S.~Ludwig, A.~Muller, and A.~Czylwik, ``Secret key generation
  from static channels with untrusted relays,'' in \emph{Proc. IEEE WiMob}, Abu
  Dhabi, United Arab Emirates, Oct. 2015, pp. 1--8.

\bibitem{9201305}
H.~M. Furqan, J.~M. Hamamreh, and H.~Arslan, ``New physical layer key
  generation dimensions: Subcarrier indices/positions-based key generation,''
  \emph{{IEEE} Commun. Lett.}, vol.~25, no.~1, pp. 59--63, 2021.

\bibitem{7458871}
X.~Zhu, F.~Xu, E.~Novak, C.~C. Tan, Q.~Li, and G.~Chen, ``Using wireless link
  dynamics to extract a secret key in vehicular scenarios,'' \emph{{IEEE}
  Trans. Mobile Comput.}, vol.~16, no.~7, pp. 2065--2078, 2017.

\bibitem{Jana2009effectiveness}
S.~Jana, S.~N. Premnath, M.~Clark, S.~K. Kasera, N.~Patwari, and S.~V.
  Krishnamurthy, ``On the effectiveness of secret key extraction from wireless
  signal strength in real environments,'' in \emph{Proc. ACM MobiCom}, Beijing,
  China, Sep. 2009, pp. 321--332.

\bibitem{PCA}
G.~Li, A.~Hu, J.~Zhang, and et~al., ``High-agreement uncorrelated secret key
  generation based on principal component analysis preprocessing,''
  \emph{{IEEE} Trans. Commun.}, vol.~66, no.~7, pp. 3022--3034, 2018.

\bibitem{USRP}
\BIBentryALTinterwordspacing
E.~research, \emph{USRP N200/N210 NETWORKED SERIES}, Mountain View, CA, 2012.
  [Online]. Available: \url{http://www.ettus.com/}
\BIBentrySTDinterwordspacing

\bibitem{Switch}
\BIBentryALTinterwordspacing
Mini-Circuits, \emph{Solid state USB RF SP8T Switch: USB-1SP8T-63H}, Brooklyn,
  NY, 2012. [Online]. Available: \url{https://www.minicircuits.com}
\BIBentrySTDinterwordspacing

\bibitem{rukhin2000}
A.~Rukhin, J.~Soto, J.~Nechvatal, M.~Smid, and E.~Barker, ``A statistical test
  suite for random and pseudorandom number generators for cryptographic
  applications,'' DTIC Document, Tech. Rep., 2001.

\bibitem{kmeans}
J.~Tou and R.~Gonzalez, \emph{Pattern Recognition Principles}.\hskip 1em plus
  0.5em minus 0.4em\relax Addison-Wesley, 1977.

\end{thebibliography}
\end{document}